\documentclass[aps,prl,reprint,superscriptaddress,floatfix]{revtex4-1}
\usepackage{graphicx} % For figures
\usepackage{color,dcolumn,bm,amssymb,amsmath} % Math and symbols
\usepackage{siunitx} % SI units
\usepackage{epstopdf}
\usepackage{hyperref} % For hyperlinks in references

\begin{document}

% Title
\title{Imaging flat band electron hydrodynamics in biased bilayer graphene}

% Authors and Affiliations
\author{Canxun Zhang}\thanks{These authors contributed equally.}
\affiliation{Department of Physics, University of California, Santa Barbara, California 93106, USA}
\author{Evgeny Redekop}\thanks{These authors contributed equally.}
\affiliation{Department of Physics, University of California, Santa Barbara, California 93106, USA}
\author{Hari Stoyanov}\thanks{These authors contributed equally.}
\affiliation{Department of Physics, University of California, Santa Barbara, California 93106, USA}
\author{Jack H. Farrell}
\affiliation{Department of Physics and Center for Theory of Quantum Matter, University of Colorado, Boulder, Colorado 80309, USA}
\author{Sunghoon Kim}
\affiliation{Department of Physics, University of California, Santa Barbara, California 93106, USA}
\author{Ludwig Holleis}
\affiliation{Department of Physics, University of California, Santa Barbara, California 93106, USA}
\author{David Gong}
\affiliation{Department of Physics, University of California, Santa Barbara, California 93106, USA}
\author{Aidan Keough}
\affiliation{Department of Physics, University of California, Santa Barbara, California 93106, USA}
\author{Youngjoon Choi}
\affiliation{Department of Physics, University of California, Santa Barbara, California 93106, USA}
\author{Takashi Taniguchi}
\affiliation{Research Center for Materials Nanoarchitectonics, National Institute for Materials Science, 1-1 Namiki, Tsukuba 305-0044, Japan}
\author{Kenji Watanabe}
\affiliation{Research Center for Electronic and Optical Materials, National Institute for Materials Science, 1-1 Namiki, Tsukuba 305-0044, Japan}
\author{Martin E. Huber}
\affiliation{Departments of Physics and Electrical Engineering, University of Colorado, Denver, Colorado 80204, USA}
\author{Ania C. Bleszynski Jayich}
\affiliation{Department of Physics, University of California, Santa Barbara, California 93106, USA}
\author{Andrew  Lucas}
\affiliation{Department of Physics and Center for Theory of Quantum Matter, University of Colorado, Boulder, Colorado 80309, USA}
\author{Andrea F. Young}
\email{andrea@physics.ucsb.edu}
\affiliation{Department of Physics, University of California, Santa Barbara, California 93106, USA}

\begin{abstract}
Hydrodynamic electron transport arises when carrier kinetics are dominated by interelectron collisions rather than the relaxation of momentum out of the electron system\cite{Gurzhi_Hydrodynamic_1968}. 
In recent years, signatures of electron hydrodynamics have been reported in graphene devices\cite{Bandurin_Negative_2016,KrishnaKumar_Superballistic_2017,Bandurin_Fluidity_2018,Berdyugin_Measuring_2019,Sulpizio_Visualizing_2019,Ku_Imaging_2020,Kim_Control_2020,Jenkins_Imaging_2022,Kumar_Imaging_2022,Huang_Electronic_2023,Krebs_Imaging_2023,Palm_Observation_2024,Talanov_Observation_2024} owing to the low disorder and weak electron--phonon coupling. 
However, these experiments have been performed in regimes where the carrier mass is light, and the electron--electron collision length---though smaller than corresponding lengths for phonon or impurity scattering---remains large in absolute terms, typically several hundred nanometers. 
This restricts hydrodynamic transport phenomena to large length scales, limiting miniaturization of devices based on hydrodynamic flow. 
The advent of dual-gated rhombohedral graphene multilayers introduces a new route toward enhanced hydrodynamic behavior via their large---and tunable---effective mass\cite{Koshino_Trigonal_2009}. 
Here, we employ a scanning superconducting magnetic sensor\cite{Finkler_Self-Aligned_2010,Vasyukov_Scanning_2013} to image local current flow in dual-gated bilayer graphene. 
Exploiting a sample geometry sensitive to both laminar and vortical flow\cite{Aharon-Steinberg_Direct_2022,Palm_Observation_2024}, we identify three distinct transport regimes---ballistic, hydrodynamic, and diffusive---across the full phase space spanned by carrier density and displacement field. 
The strongest hydrodynamic transport is observed in the flat band regime, where fitting our results to a unified Boltzmann transport model reveals the electron--electron scattering length to be comparable to the Fermi wavelength of $\sim\qty{50}{nm}$. 
High-current measurements, meanwhile, reveal striking nonlinearities in the flow pattern. 
Our results pave the way for miniaturized electronic devices based on linear and nonlinear electron hydrodynamics. 
\end{abstract}

\maketitle

Hydrodynamics is a universal framework that describes how conserved quantities such as momentum evolve toward local equilibrium in thermalizing many-body systems\cite{Lucas_Hydrodynamics_2018,Fritz_Hydordynamic_2024}. 
Electrons in solids typically equilibrate with each other via momentum-conserving electron--electron scattering, parameterized by a length scale, $\ell_\mathrm{ee}$\cite{Gurzhi_Hydrodynamic_1968}. 
For hydrodynamic flow of electrons to be observed, $\ell_\mathrm{ee}$ must be the shortest scattering length at play; in particular it must be smaller than both the device size $L$ and $\ell_\mathrm{mr}$, the length scale associated with relaxation of momentum to an external bath. 
In ordinary metals, $\ell_\mathrm{mr}\ll\ell_\mathrm{ee},L$ due to scattering from phonons and disorder as well as Umklapp processes, leading to diffusive transport. 
In graphene\cite{Narozhny_Hydrodynamics_2015,Levitov_Electron_2016}, the absence of two-body Umklapp scattering, weak electron--phonon coupling, and continuous improvements in material quality now allow for $\ell_\mathrm{mr}$ to far exceed the typical device size of $L\approx\qty{10}{\um}$ over a wide range of temperatures. 
Past experiments have shown that, while at low temperatures transport is ballistic (i.e., $\ell_\mathrm{mr},\ell_\mathrm{ee}\gg L$), increasing the temperature can lower $\ell_\mathrm{ee}$ sufficiently to enable the observation of hydrodynamic effects\cite{Bandurin_Negative_2016,KrishnaKumar_Superballistic_2017,Bandurin_Fluidity_2018,Berdyugin_Measuring_2019,Sulpizio_Visualizing_2019,Ku_Imaging_2020,Kim_Control_2020,Jenkins_Imaging_2022,Kumar_Imaging_2022,Huang_Electronic_2023,Krebs_Imaging_2023,Palm_Observation_2024,Talanov_Observation_2024}. 
Quantitatively, however, these experiments found $\ell_\mathrm{ee}\gtrsim\qty{250}{nm}$, limited by the eventual onset of phonon scattering. 
This precludes further device miniaturization and restricts the range over which emergent hydrodynamic parameters, such as viscosity, can be tuned in experiments. 

\begin{figure}[ht!]
\centering
\includegraphics[width=2.5in]{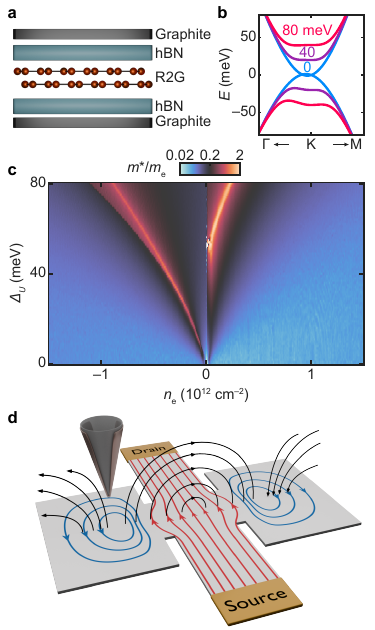}
\caption{\textbf{Tunable effective mass in dual-gated R2G.}
\textbf{a}, Schematic view of a dual-gated R2G device. Two graphite gates, separated from the R2G flake by hexagonal boron nitride (hBN) dielectrics, together control the carrier density $n_\mathrm{e}$ and the applied displacement field $D$. 
\textbf{b}, Single-particle band structure of R2G near the Brillouin zone corner calculated within a tight-binding model\cite{Jung_Accurate_2014}. Different curves correspond to different values of interlayer potential difference $\Delta_U\propto D$. 
\textbf{c}, Density of states effective mass $m^*$ as a function of $n_\mathrm{e}$ and $\Delta_U$. 
\textbf{d}, Illustration of the experimental geometry, showing nSOT sensor, current streamlines (red and blue), and flux lines associated with the fringe magnetic field (black). 
}
\label{fig:fig1}
\end{figure}

In a two-dimensional Fermi liquid with parabolic dispersion, 
\begin{align}
    \ell_\mathrm{ee}\propto \frac{\left|n_\mathrm{e}\right|^{3/2}}{m^{*2} T^2},
    \label{eq1}
\end{align}
where $n_\mathrm{e}$ is the carrier density, $m^*$ is the effective mass, and $T$ is the temperature\cite{Gurzhi_Hydrodynamic_1968,Hui_Hydrodynamics_2025} (Methods). 
The large $\ell_\mathrm{ee}$ characterizing past experiments in graphene can be tied to the fact that $m^*$ is small. 
In rhombohedral (ABC-stacked) multilayer graphene, including AB-stacked bilayer (R2G; Fig.~\ref{fig:fig1}a), the electronic structure features two low-energy bands with an effective mass that can be tuned by over one order of magnitude as a function of $n_\mathrm{e}$ and displacement field $D$\cite{Koshino_Trigonal_2009}, both of which can be controlled by the electric field effect in dual-gated devices. 
At $D=0$ (the regime of past experiments\cite{Bandurin_Negative_2016,Berdyugin_Measuring_2019}), the band structure is semimetallic, with $m^*\approx0.05m_\mathrm{e}$ where $m_\mathrm{e}$ is the free electron mass (Fig.~\ref{fig:fig1}b,c). 
Ballistic transport is thus expected at low temperatures, except near the charge neutrality point (CNP), where transport is instead dominated by electron--hole recombination. 
At large $D$, in contrast, the interlayer potential difference $\Delta_U$ produces both a band gap and band edge van Hove singularities which increase $m^*$ to values comparable to or even larger than $m_\mathrm{e}$. 
Per Eq.~\eqref{eq1}, these flat band regions should be highly favorable to hydrodynamic transport. 

Complicating this naive picture, however, is the onset of Fermi surface reconstructions due to Stoner ferromagnetism at large $m^*$. 
Experiments in dual-gated R2G have revealed a complex low-temperature phase diagram of correlated ferromagnetic phases with broken spin and/or valley symmetries\cite{Zhou_Isospin_2022,delaBarrera_Cascade_2022,Seiler_Divergent_2022,Zhang_Enhanced_2023,Li_Tunable_2024,Holleis_Nematicity_2025,Zhang_Twist-programmable_2025}. 
Notably, the resistance in the metallic high-temperature parent phase of these magnets was observed to show a negative temperature coefficient, $\mathrm{d}R/\mathrm{d}T<0$\cite{Holleis_Fluctuating_2025,Seiler_Interaction-Driven_2024}. 
This behavior is reminiscent of the hydrodynamic Gurzhi effect\cite{Gurzhi_Hydrodynamic_1968} where the declining viscosity of the electron fluid leads to a negative temperature coefficient\cite{spivak,lucashartnoll}. 
However, negative temperature coefficients have also been observed near the Curie temperature in elemental rare-earth magnets\cite{Colvin_Electrical_1960}, where this behavior is attributed to reduced $\ell_\mathrm{mr}$ due to scattering of electrons by magnetic fluctuations that subsequently relax their momentum via Umklapp processes\cite{Fisher_Resistive_1968}. 
While these mechanisms produce similar behavior of the global resistivity, they are expected to generate qualitatively distinct microscopic current profiles. 

Here, we use a superconducting sensor\cite{Finkler_Self-Aligned_2010,Vasyukov_Scanning_2013} to measure the fringe magnetic field above a dual-gated R2G device, which we then use to reconstruct the in-plane current profile on the nanosccale as a function of $n_\mathrm{e}$ and $D$ (Methods). 
A key challenge in spatially resolved measurements of this kind is to successfully disambiguate all three (hydrodynamic, ballistic, and diffusive) transport regimes. 
Our sample geometry is shown in Fig.~\ref{fig:fig1}d, and features both a fixed-width channel ($w\approx\qty{1.15}{\um}$) and two lateral chambers connected to the channel by constrictions of $d\approx\qty{1.0}{\um}$ (Methods, Extended Data Fig.~\ref{fig:ED_sample}). 
Electron transport in the channel distinguishes hydrodynamic flow, characterized by a Poiseuille velocity profile with a higher current density in the center, from ballistic or diffusive behavior with flatter or even concave profiles\cite{Gurzhi_Hydrodynamic_1968,Torre_Nonlocal_2015,Sulpizio_Visualizing_2019,Ku_Imaging_2020,Huang_Electronic_2023}. 
In contrast, current flow in the lateral chambers probes the local violation of Ohm's law---stationary vortices or whirlpools can emerge in either the ballistic or hydrodynamic regime\cite{Torre_Nonlocal_2015,Levitov_Electron_2016,Bandurin_Negative_2016,Bandurin_Fluidity_2018,Aharon-Steinberg_Direct_2022,Palm_Observation_2024} but are forbidden when transport is diffusive on short length scales. 
Cross-correlating flow patterns in the channel and chambers can thus be used to unambiguously determine the nature of the carrier kinetics. 

\section{Imaging local current flow}

\begin{figure*}[ht!]
\centering
\includegraphics[width=7in]{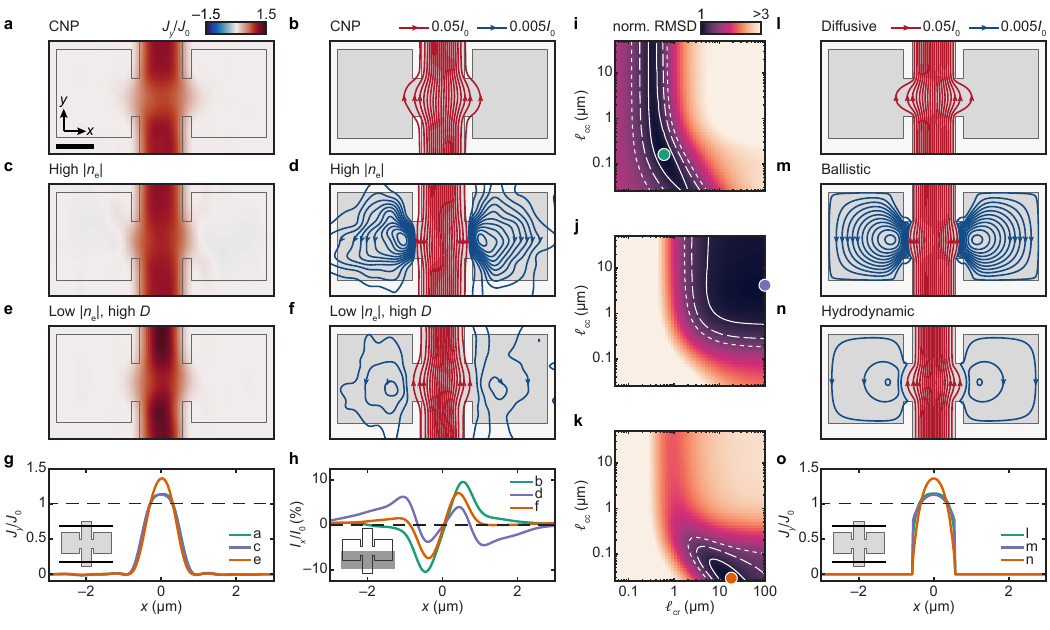}
\caption{\textbf{Contrasting microscopic transport regimes.}
\textbf{a}, Reconstructed current density $J_y(x,y)$ at $n_\mathrm{e}=\qty{0e12}{cm^{-2}}$ and $D=\qty{0}{V/nm}$, where R2G is a charge-neutral gapless semimetal. Scale bar: \qty{1}{\um}. $J_y$ is normalized by $J_0\equiv I_0/w$, the average current density in the channel. 
\textbf{b}, Current streamlines at the same $n_\mathrm{e}$ and $D$ as panel a. 
\textbf{c}, $J_y(x,y)$ for $n_\mathrm{e}=\qty{-1.3e12}{cm^{-2}}$ and $D=\qty{0.39}{V/nm}$. 
\textbf{d}, Current streamlines at the same $n_\mathrm{e}$ and $D$ as panel c. 
\textbf{e}, $J_y(x,y)$ for $n_\mathrm{e}=\qty{-0.4e12}{cm^{-2}}$ and $D=\qty{0.39}{V/nm}$. 
\textbf{f}, Current streamlines at the same $n_\mathrm{e}$ and $D$ as panel e. 
\textbf{g}, $J_y(x)/J_0$ averaged between measurements at $y=\pm\qty{2}{\um}$ (see inset) for the data in panels a, c, and e as marked. 
\textbf{h}, Normalized transverse current $I_x(x)/I_0$ (see inset and Eq.~\eqref{eq2} of main text) corresponding to the data in panels b, d, and f as marked. 
\textbf{i}, Combined root mean square deviation (RMSD) between experimental data at $n_\mathrm{e}=\qty{0e12}{cm^{-2}}$ and $D=\qty{0}{V/nm}$ and the linearized Boltzmann model parameterized by $\ell_\mathrm{cr}$ and $\ell_\mathrm{cc}$. The RMSD is normalized by the minimum value, indicated by the dot and corresponding to the best fit. Contours of 1.1, 1.3, and 1.5 delineate phenomenological thresholds in the RMSD. 
\textbf{j}, RMSD for $n_\mathrm{e}=\qty{-1.3e12}{cm^{-2}}$ and $D=\qty{0.39}{V/nm}$. 
\textbf{k}, RMSD for $n_\mathrm{e}=\qty{-0.4e12}{cm^{-2}}$ and $D=\qty{0.39}{V/nm}$. 
\textbf{l}, Theoretical current streamlines calculated from the Boltzmann model for $\ell_\mathrm{cr}=\qty{600}{nm}$ and $\ell_\mathrm{cc}=\qty{161}{nm}$, corresponding to the best fit in panel i. 
\textbf{m}, Theoretical current streamlines for $\ell_\mathrm{cr}=\qty{100}{\um}$ and $\ell_\mathrm{cc}=\qty{4.2}{\um}$, corresponding to the best fit in panel j. 
\textbf{n}, Theoretical current streamlines for $\ell_\mathrm{cr}=\qty{18.2}{\um}$ and $\ell_\mathrm{cc}=\qty{29}{nm}$, corresponding to the best fit in panel k. 
\textbf{o}, $J_y(x)/J_0$ at $y=\pm\qty{2}{\um}$ for the best fits described in panels l, m and n. 
}
\label{fig:fig2}
\end{figure*}

Figure~\ref{fig:fig2}a shows a spatial map of the current density along the channel direction, $J_y(x,y)$, for a total applied current of $I_0\approx\qty{4.5}{\uA}$ at the CNP and the value of $D$ corresponding to $\Delta_U=0$. 
$J_y(x,y)$ is determined from the measured $B$ field (shown in Extended Data Fig.~\ref{fig:ED_B_J_stream_fig2} along with spatial derivatives and the corresponding $J_x(x,y)$) by standard Fourier domain techniques (Methods). 
A complementary visualization is provided by the streamline plot shown in Fig.~\ref{fig:fig2}b, in which each red line represents $I_0/20$. 
At the CNP, streamlines in the center of the channel follow a straight path, but those at the left and right extremes take a slight detour into the chambers. 
Each streamline enters the scan frame from the bottom and exits it at the top without forming any closed loop, in accordance with local Ohm's law $\mathbf{J}=-\sigma\nabla\phi$ where $\sigma$ is the local conductivity and $\phi$ is the local electric potential. 
A different picture emerges under heavy electron or hole doping, shown in Fig.~\ref{fig:fig2}c,d for $n_\mathrm{e}=\qty{-1.3e12}{cm^{-2}}$ and $D=\qty{0.39}{V/nm}$. 
Here, $J_y$ in the channel resembles that at the CNP, but the streamline plot reveals closed loops of current within the chambers (here each blue line represents $I_0/200$). 
A third qualitatively distinct regime is found at high $D$ and low $|n_\mathrm{e}|$, corresponding to the flat band regime. 
Figure~\ref{fig:fig2}e,f show $J_y$ and the streamlines for $n_\mathrm{e}=\qty{-0.4e12}{cm^{-2}}$ and $D=\qty{0.39}{V/nm}$. 
Within the channel, $J_y$ is significantly concentrated away from the boundaries, while vortical flow is still present in the chambers, albeit with quantitatively distinct strength and spatial distribution. 

We define two empirical metrics that capture both the laminar channel flow and vortical chamber flow. 
The channel flow profile is characterized by $J_y(x)$ measured far from the entrance to lateral chambers (near $y=\pm\qty{2}{\um}$). 
As shown in Fig.~\ref{fig:fig2}g, the flat band regime features a current profile with approximately \qty{30}{\percent} higher concentration than the average current density $J_0$ (dashed line). 
This contrasts with the much flatter distribution observed for both the CNP and high-density regimes. 
To quantify the vortex profile, we plot the the ``transverse current'' $I_x$ (Fig.~\ref{fig:fig2}h), defined as 
\begin{align}
    I_x(x):=\frac{\int_{-y_0}^0J_x(x,y)\,\mathrm{d}y-\int_0^{y_0}J_x(x,y)\,\mathrm{d}y}{2},
    \label{eq2}
\end{align}
where $y_0=\qty{2}{\um}$. 
Physically, $I_x$ counts the total current (i.e., number of streamlines) passing through a vertical cross-section at position $x$. 
At the CNP, where no vorticity is expected or observed in the streamlines, $I_x$ is positive for positive $x$ and negative for negative $x$. 
Vortical flow, in contrast, introduces sign changes in $I_x$. 
For vortices on the central horizontal axis, the magnitude of this sign-changing signal approximates their strength while the position of the extrema is closely correlated with their center. 
Evidently, between the high-$|n_\mathrm{e}|$ regime of Fig.~\ref{fig:fig2}c,d and the flat band regime of Fig.~\ref{fig:fig2}e,f, the vortical flow decreases in strength while the vortex center moves further into the chambers. 

To quantify these behaviors, we compare experimental data with a phenomenological linearized Boltzmann transport model applied to the experimental sample geometry, in which scattering is parameterized via a two-rate relaxation time approximation (Methods). 
The free parameters in the model are the current relaxing scattering length, $\ell_\mathrm{cr}$, and the current-conserving scattering length, $\ell_\mathrm{cc}$, as well as the boundary conditions. 
$\ell_\mathrm{cr}$ includes microscopic processes such as electron--impurity, electron--phonon, and electron--hole scatterings that relax the electric current. 
$\ell_\mathrm{cc}$ includes scattering between carriers of the same polarity which does not reduce the total current. 
We find that fits are only weakly sensitive to the boundary conditions as long as they are at least partially diffusive (see Extended Data Fig.~\ref{fig:ED_boundary}), and so fix the boundary as fully diffusive. 
We fit our data by calculating root mean square deviations (RMSDs) between experiment and theory that capture the contrasts in the $J_y/J_0$ and $I_x/I_0$ curves, and combining these RMSDs into a single figure of merit which we then normalize by the minimum value for a given data set (Methods). 
This combined RMSD is plotted as a function of $\ell_\mathrm{cr}$ and $\ell_\mathrm{cc}$ in Fig.~\ref{fig:fig2}i--k for the data of panels g and h. 
These figures also show contours defining phenomenological thresholds corresponding to where the RMSD exceeds the minimum by \qty{10}{\percent}, \qty{30}{\percent}, and \qty{50}{\percent}. 
We note that the ``goodness of fit'' implied by these thresholds is dominated by systematic errors arising from imprecise knowledge of the device geometry or mesoscopic asymmetries not captured in our model, rather than random noise in our magnetometry measurement. 
Comparison of data with model outputs is shown in Extended Data Fig.~\ref{fig:ED_fit_Jy_Ix}. 

At the CNP (Fig.~\ref{fig:fig2}i), the best fit is achieved for $\ell_\mathrm{cr}$ on the order of a few hundred nanometers while $\ell_\mathrm{cc}$ is nearly unconstrained. 
This is consistent with diffusive transport phenomenology, with $\ell_\mathrm{cr}$ agreeing with values obtained from mobility measurement on a similar device and prior literature\cite{Wang_One-Dimensional_2013}, $\ell_\mathrm{cr}\sim\qty{1}{\um}$. 
We note that microscopically, charge neutral, gapless graphene is better described as an electron--hole plasma\cite{Crossno_Observation_2016,Tan_Dissipation-enabled_2022}. 
This effect is not captured by our model, but is also not expected to manifest strongly in the flow profile of \emph{electric} current. 
Indeed, we find that our single-band Boltzmann model, which captures the fact that current relaxes on small length scales, provides a good fit to the experimental data. 
At high $|n_\mathrm{e}|$ (Fig.~\ref{fig:fig2}j), the best fit is obtained for simultaneously large $\ell_\mathrm{cr}$ and $\ell_\mathrm{cc}$, i.e., this regime is ballistic and the strong vortical flow arises from ballistic carrier trajectories scattering off the chamber boundaries. 
Finally, in the high-$D$, low-$|n_\mathrm{e}|$ regime (Fig.~\ref{fig:fig2}k), the Boltzmann fit finds a large $\ell_\mathrm{cr}\gtrsim L$ while $\ell_\mathrm{cc}$ is as small as \qty{50}{nm}. 
This large separation in scattering lengths implies that the flow is deep in the hydrodynamic regime, where current-conserving collisions dominate transport and the current pattern is determined by the Gurzhi length, $\sqrt{\ell_\mathrm{cr}\ell_\mathrm{cc}}/2\approx\qty{360}{nm}$. 
The assignment of these three flow regimes is further supported by direct visualization of the calculated current distributions at the representative best-fit parameters (dots in Fig.~\ref{fig:fig2}i--k). 
The streamline plots of Fig.~\ref{fig:fig2}l--n exhibit strong/weak vortical flow in the ballistic/hydrodynamic regime and an absence of vortical flow in the diffusive regime, in qualitative agreement with the experimental maps of Fig.~\ref{fig:fig2}b,d,f. 
Calculated $J_y(x)/J_0$ (Fig.~\ref{fig:fig2}o) similarly agrees with the Poiseuille flow in the channel for the hydrodynamics regime. 
We note that the unphysical current streamlines which exit the device area in the experimental data arise due to the finite spatial resolution of the nSOT probe as well as numerical artifacts associated with Fourier-domain inversion. 
Our quantitative fitting is performed by solving the forward problem---i.e., generating expected experimental signal from the microscopic current distribution (see Methods and Extended Data Fig.~\ref{fig:ED_sim_vs_rec} for more details). 

\section{Flow regime phase diagram}

\begin{figure*}[ht!]
\centering
\includegraphics[width=7in]{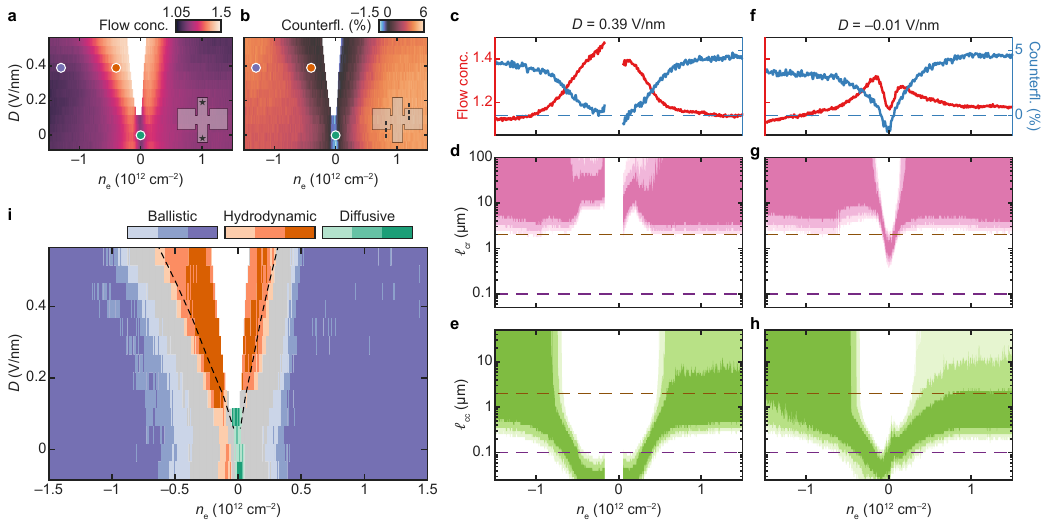}
\caption{\textbf{Transport regime phase diagram.}
\textbf{a}, Flow concentration $J_y(x=0)/J_0$ and 
\textbf{b}, Counterflow $I_x(|x|=\qty{1.3}{\um})/I_0$ (see main text) as a function of $n_\mathrm{e}$ and $D$. Dots represent parameters of measurements presented in Fig.~\ref{fig:fig2}. 
\textbf{c}, Flow concentration and counterflow as a function of $n_\mathrm{e}$ at $D=\qty{0.39}{V/nm}$. 
\textbf{d}, Fitting ranges for $\ell_\mathrm{cr}$ and 
\textbf{e}, $\ell_\mathrm{cc}$ for the data in panel c. 
The dark, intermediate, and light shaded regions in panels d and e denote ranges where the normalized RMSD is less than 1.1, 1.2, and 1.3, respectively. Dashed lines denote \qty{100}{nm} and \qty{2}{\um}. 
\textbf{f}, Flow concentration and counterflow as a function of $n_\mathrm{e}$ at $D=\qty{-0.01}{V/nm}$. 
\textbf{g}, Fitting ranges for $\ell_\mathrm{cr}$ and 
\textbf{h}, $\ell_\mathrm{cc}$ for the data in panel f. 
\textbf{i}, Combined phase diagram indicating transport regimes. Blue represents the ballistic regime where $\min{\ell_\mathrm{cr}}>\qty{2}{\um}$ and $\min{\ell_\mathrm{cc}}>\qty{100}{nm}$. Orange represents the hydrodynamic regime where $\min{\ell_\mathrm{cr}}>\qty{2}{\um}$ and $\max{\ell_\mathrm{cc}}<\qty{100}{nm}$. Green represents the diffusive regime where $\max{\ell_\mathrm{cr}}<\qty{2}{\um}$. Different shadings correspond to thresholds of normalized RMSD~$<1.1,1.2,1.3$ and gray regions indicate regime crossovers. Dashed lines delimit the extent of the regime of fluctuating magnetic moments and negative $\mathrm{d}R/\mathrm{d}T$ reported in \cite{Holleis_Fluctuating_2025}. 
}
\label{fig:fig3}
\end{figure*}

The quantitative analysis developed for the three $(n_\mathrm{e},D)$ points described in Fig.~\ref{fig:fig2} can be extended across the full parameter space to map the domains of ballistic, hydrodynamic, and diffusive transport. 
At a qualitative level, the evolution of flow patterns can be characterized by two further simplified quantities. 
Flow concentration is defined by $J_y(x=0)/J_0$, i.e., the enhancement of current flow at the channel center. 
The ``counterflow~\%'', meanwhile, is defined by $I_x$ evaluated at $|x|=\qty{1.3}{\um}$ and antisymmetrized between the two chambers; this approximates the total loop current near the vortex center. 
Figure~\ref{fig:fig3}a,b show the flow concentration and counterflow~\% as a function of $n_\mathrm{e}$ and $D$. 
Evidently, the three representative points of Fig.~\ref{fig:fig2} lie within broad domains of stability, with high $|n_\mathrm{e}|$ characterized by strong counterflow and low flow concentration, lower $|n_\mathrm{e}|$ and high $D$ characterized by weak counterflow and high concentration, and a small regime at the CNP and low $D$ showing neither positive counterflow nor high concentration. 

Fitting data taken across the entire phase space to the Boltzmann model allows us to directly constrain the transport length scales. 
For example, at high $D$ the qualitative crossover from low flow concentration/high counterflow to high concentration/low counterflow (Fig.~\ref{fig:fig3}c) can be linked purely to a collapse in the value of $\ell_\mathrm{cc}$ at low density while $\ell_\mathrm{cr}$ remains larger than the device size (Fig.~\ref{fig:fig3}d,e). 
In other words, lowering the density leads to a ballistic-to-viscous crossover. 
At $D\approx0$, in contrast (Fig.~\ref{fig:fig3}f--h), three regimes are visible as a function of $n_\mathrm{e}$, with ballistic transport at high $|n_\mathrm{e}|$ separated from a narrow region of diffusive transport near the CNP by at least one regime of viscous transport at low hole density. 

We note that transport regimes are not sharply bounded, with transitions instead being described by crossover-like behavior. 
Thus, even though we clearly observe deeply ballistic, hydrodynamic, and diffusive regimes, any definition of their boundaries is subject to some degree of arbitrariness. 
Based on the characteristic device size of $w\sim d\sim\qty{1}{\um}$, we define two thresholds of \qty{100}{nm} and \qty{2}{\um} (marked in Fig.~\ref{fig:fig3}d,e,g,h; see Extended Data Fig.~\ref{fig:ED_alt_phase} for more details). 
The ballistic regime is defined as where a given RMSD threshold condition is satisfied for $\ell_\mathrm{cr}>\qty{2}{\um}$ and $\ell_\mathrm{cc}>\qty{100}{nm}$. 
The hydrodynamic regime is defined similarly, for $\ell_\mathrm{cr}>\qty{2}{\um}$ and $\ell_\mathrm{cc}<\qty{100}{nm}$, while for the diffusive regime $\ell_\mathrm{cr}<\qty{2}{\um}$. 
Applying these criteria to the experimentally determined $\ell_\mathrm{cr}$ and $\ell_\mathrm{cc}$ yields ballistic, hydrodynamic, and diffusive regions shown in Fig.~\ref{fig:fig3}i (here darker shading indicates regions of higher confidence based on the RMSD from the model fit). 
Regions rendered in gray indicate crossovers.  

The hydrodynamic regime coincides with the large-$m^*$ region of Fig.~\ref{fig:fig1}c. 
Over much of this regime, our semiclassical Boltzmann fit gives $\ell_\mathrm{cc}\lesssim\qty{50}{nm}$, comparable to the Fermi wavelength $\lambda_\mathrm{F}=2\sqrt{\pi/|n_\mathrm{e}|}$ at low densities; in other words, electrons scatter off of each other on a length scale comparable to their separation. 
This is the outer limit of the validity of the semiclassical approximation that underlies Boltzmann transport, and implies that transport is hydrodynamic on the smallest allowable length scales. 
Phenomenologically, our finding also implies that the negative $\mathrm{d}R/\mathrm{d}T$ observed in bulk transport measurements of rhombohedral multilayers\cite{Holleis_Fluctuating_2025,Seiler_Interaction-Driven_2024} indeed arises from the Gurzhi effect, with the temperature-dependent viscosity dominating the resistance behavior for a finite-width channel. 
One implication of this finding is that if samples can be fabricated with dimensions considerably larger than the Gurzhi length, this resistance effect should weaken. 

\begin{figure*}[ht!]
\centering
\includegraphics[width=7in]{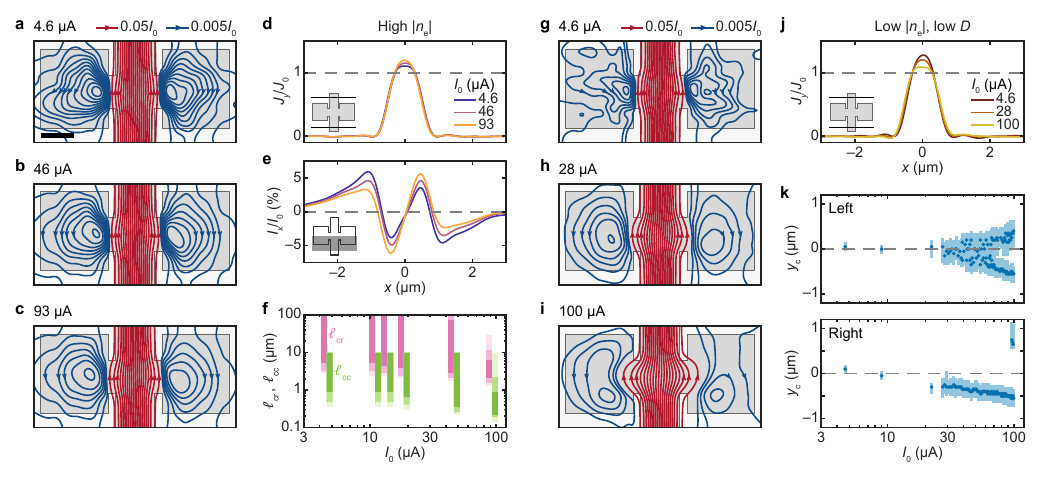}
\caption{\textbf{Visualizing current-driven nonlinearities.}
\textbf{a}, Current streamlines for $n_\mathrm{e}=\qty{-1.5e12}{cm^{-2}}$ and $D=\qty{0.39}{V/nm}$ for $I_0=\qty{4.6}{\uA}$,
\textbf{b}, $I_0=\qty{46}{\uA}$, and 
\textbf{c}, $I_0=\qty{93}{\uA}$. Scale bar: \qty{1}{\um}. 
\textbf{d}, $J_y(x,\pm\qty{2}{\um})/J_0$ for the data in panels a, b and c. 
\textbf{e}, $I_x(x)/I_0$ for the data in panels a, b and c. 
\textbf{f}, Fitting ranges for $\ell_\mathrm{cr}$ and $\ell_\mathrm{cc}$ as a function of $I_0$. Shadings represent ranges where normalized RMSD~$<1.1,1.2,1.3$. 
\textbf{g}, Current streamlines for $n_\mathrm{e}=\qty{-0.25e12}{cm^{-2}}$ and $D=\qty{0}{V/nm}$ for $I_0=\qty{4.6}{\uA}$, 
\textbf{h}, $I_0=\qty{28}{\uA}$, and 
\textbf{i}, $I_0=\qty{100}{\uA}$. 
\textbf{j}, $J_y(x,\pm\qty{2}{\um})/J_0$ for the data in panels g, h and i. 
\textbf{k}, The vortex center position(s), $y_\mathrm{c}$, for the left and right chambers as a function of $I_0$ (see Methods). $y=0$ denotes the horizontal mirror axis of the device. 
}
\label{fig:fig4}
\end{figure*}

Previous thermodynamic measurements at similar values of $T$, $n_\mathrm{e}$ and $D$ have revealed a ``local moment'' entropy associated with isospin fluctuations\cite{Holleis_Fluctuating_2025}; the magnitude of this entropy implies an isospin correlation length that is also comparable to $\lambda_\mathrm{F}$. 
Gate-biased bilayer graphene (and presumably other rhombohedral multilayers) is thus in a nearly classical liquid state in which both the electron--electron scattering and isospin correlations are as short-range as possible given the particle density. 
Within this picture, our observation of hydrodynamics implies that electrons scatter strongly with these moments, which then in turn return momentum to electron system at a faster rate than they dissipate it to the lattice. 
The short $\ell_\mathrm{cc}$ implies that the state is above the coherence temperature of the Fermi liquid. 
At the same time, our $T\approx\qty{2}{K}$ is well below the bare energy scales set by the Fermi and interaction energies, which are both in the few-\unit{meV} range; these determine the effective Debye temperature for the electrons, defined as the frequency with which particles vibrate in the cage formed by their neighbors. 
The phase studied here may thus microscopically be a ``semiquantum liquid''\cite{Andreev_Kinetic_1979} in which the carrier dynamics are dominated by configurational changes between states with solid-like short-range correlations. 

\section{Imaging nonlinear transport}

Our superconducting sensor puts an upper bound on the equilibrium temperature of the experimental setup, and all of the data presented above are acquired at a cryostat temperature of \qty{1.7}{K}. 
However, it is well known that the weakness of electron--phonon coupling in graphene allows the electron temperature to be much higher than the lattice temperature, particularly under large current drive\cite{Meric_Current_2008,Massicotte_Hot_2021,Srivastav_Revealing_2025}. 
In our imaging experiment, the onset of such a nonlinearity is expected to manifest as a change in the microscopic current distribution. 
Indeed, for the analysis of the \qty{1.7}{K} transport regimes of Figs.~\ref{fig:fig2},~\ref{fig:fig3}, we take pains to measure with sufficiently small current that the flow profile is observed to be independent of the current amplitude (see Extended Data Fig.~\ref{fig:ED_linearity} for details). 
More generally, however, heating of the electron system is expected to strongly favor hydrodynamic transport, as it reduces $\ell_\mathrm{cc}$ but may have a diminished effect on $\ell_\mathrm{cr}$ if the phonon bath remains cold. 
This effect is exemplified in the current-driven evolution of the flow profile at $n_\mathrm{e}=\qty{-1.5e12}{cm^{-2}}$ and $D=\qty{0.39}{V/nm}$ where transport is deep in the ballistic regime at the lowest applied currents. 
As $I_0$ grows from 4.6 to \qty{93}{\uA} (Fig.~\ref{fig:fig4}a--c), the channel current profile $J_y(x)/J_0$ increases in concentration (Fig.~\ref{fig:fig4}d), whereas $I_x(x)/I_0$ in the chambers gradually decreases (Fig.~\ref{fig:fig4}e). 
Current streamline plots show sparser loops for higher $I_0$, indicating weakening of the vortical flow. 
Fitting $J_y(x)$ and $I_x(x)$ curves to the linearized Boltzmann model (Fig.~\ref{fig:fig4}f) reveals a gradually decreasing $\ell_\mathrm{cc}$, from $\ell_\mathrm{cc}\gtrsim\qty{1}{\um}$ to $\ell_\mathrm{cc}\approx\qty{500}{nm}$, while $\ell_\mathrm{cr}$ remains larger than the device size (see also Extended Data Fig.~\ref{fig:ED_fit_Fig4}). 
This incipient ballistic-to-viscous crossover resembles previous experiments in light-electron graphene systems that rely on elevated equilibrium temperatures to achieve hydrodynamic regimes\cite{Bandurin_Negative_2016,Berdyugin_Measuring_2019}, suggesting a quasi-equilibrium picture in which the temperature of the electron gas is uniformly raised due to the dissipation of the applied current. 

We also observe high-current effects that go beyond the phenomenological Boltzmann model. 
This is demonstrated in Fig.~\ref{fig:fig4}g--i, where we plot current streamlines at $n_\mathrm{e}=\qty{-0.25e12}{cm^{-2}}$ and $D=\qty{0}{V/nm}$ for several values of $I_0$. 
Here, the low-current transport is within the ballistic-to-viscous crossover regime, showing both elevated flow concentration in the channel and vortical flow in the chambers. 
Increasing $I_0$ to \qty{100}{\uA} causes a reduction in the channel flow concentration (Fig.~\ref{fig:fig4}j), suggesting a possible crossover to diffusive transport. 
This is consistent with some heating of the phonon bath in addition to an elevated electron temperature. 
However, the spatial distribution of current is qualitatively inconsistent with the Boltzmann model. 
Most strikingly, the vortex centers are driven away from the central horizontal axis of the device (Fig.~\ref{fig:fig4}k). 
This is accompanied by a gradual penetration of the central channel flow through the constrictions and into the chambers, which drives the vortical flow into a crescent shape (see Supplementary Video~\ref{video1}). 

Throughout the available parameter range in the linearized Boltzmann model spanned by $\ell_\mathrm{cr}$ and $\ell_\mathrm{cc}$, vortices are always quasi-circular and on the central axis (see Extended Data Fig.~\ref{fig:ED_center}), contrasting with the observations at high current. 
The breakdown of the model may have several origins. 
One naive possibility is a nonlinear hydrodynamic effect associated with large current flow. 
Within a purely hydrodynamic theory, the emergence of nonlinearities such as turbulence is parameterized by the Reynolds number $\mathrm{Re}=\frac{4v}{v_\mathrm{F}}\frac{L}{\ell_\mathrm{cc}}$, where $v$ is the drift velocity and $v_\mathrm{F}$ is the Fermi velocity. 
Taking an effective mass of $m^*\approx0.05m_\mathrm{e}$ at $\Delta_U=0$, this gives $\mathrm{Re}\approx40$--50 for the highest currents---far short of the onset of turbulence in typical fluid flows. 
Alternatively, the phenomenological nature of the Boltzmann model may ignore too many features of the intricate low-energy band structure of R2G to be quantitatively accurate, which may manifest most strongly at high currents. 
In particular, a large current drive may lead to spatially dependent shifts in the chemical potential which may alter the flow pattern. 
Both the density of states and the anisotropy of the Fermi surfaces are chemical potential dependent in R2G but these effects are ignored in our Boltzmann model where the band structure is assumed to be that of a single parabolically dispersing band. 
Finally, the equilibrium temperature profile (for either the electrons, phonons, or both) can be spatially nonuniform, allowing a higher temperature in the channel as compared to the chambers. 

Microscopic theoretical modeling may elucidate the origin of these effects, and help harness them for functional device applications based on nonlinear viscous transport\cite{Geurs_Rectification_2020}. 
More broadly, effective mass tuning of $\ell_\mathrm{ee}$ into a deeply hydrodynamic regime opens a new avenue to pursuing extreme hydrodynamic effects in electron fluids. 
For example, future experiments in rhombohedral multilayers of higher layer number (where $m^*$ can be even larger than in R2G) combined with custom geometries may access regimes of Reynolds number compatible with turbulent or preturbulent flow\cite{Gabbana_Prospects_2018}. 
The highly anisotropic Fermi surfaces in these devices may also facilitate the observation of anisotropic viscosity in the hydrodynamic regime\cite{PhysRevB.99.235148,Varnavides_Electron_2020}, a unique feature of electron hydrodynamics which has not been explored experimentally. 
Finally, the scheme for mapping microscopic transport regimes may shed light on the wealth of correlated electron physics now available in ultraclean crystalline two-dimensional material systems.

% Methods
\section*{Methods}

\subsection*{Sample fabrication}

Dual-gated R2G was fabricated through van der Waals stacking of bilayer graphene, hexagonal boron nitride (hBN) dielectric layers, and few-layer graphite gate plates. 
The flakes were prepared by mechanical exfoliation on Si/SiO$_2$ substrates. 
To begin the stacking process, graphite exfoliated on polydimethylsiloxane (PDMS) was dropped onto hBN. Poly(bisphenol A carbonate) (PC) film was then used to pick up the graphite-hBN pair as well as subsequent graphene, hBN, and graphite layers. 
The finished stack was dropped onto an Si/SiO$_2$ substrate, and the PC film was washed off with N-methylpyrrolidone.

Electron-beam lithography (EBL) was used to define a contact electrode pattern in a PMMA A8 950K resist mask. 
A CHF$_3$/O$_2$ ICP-RIE plasma etch followed by electron-beam deposition of chromium/palladium/gold was used to make edge contacts to individual layers of the stack. 
Then, an etch mask was defined via EBL of a PMMA A4 495K/A2 950K bilayer resist followed by electron-beam evaporation of aluminum. 
The rectangular lateral chambers were defined using a second, separate aluminum mask in order to avoid EBL proximity effects that would distort the intended device geometry. 
All EBL processes were performed using a scanning electron microscope equipped with a Nanometer Pattern Generation System. 
After the deposition of both aluminum masks, the stack was etched using a CHF$_3$/O$_2$ ICP-RIE plasma etch. 
Finally, the sample was immersed in an AZ300MIF solution for 20 minutes to remove the aluminum.

\subsection*{nSOT measurements}

Local magnetometry measurements were performed in a liquid helium cryostat at a base temperature of \qty{1.7}{K} using a superconducting quantum interference device on the apex of a quartz tube (nanoSQUID-on-tip, nSOT). 
The probe was fabricated following established protocols\cite{Finkler_Self-Aligned_2010,Vasyukov_Scanning_2013,Redekop_Direct_2024}. 
A quartz micropipette with an inner diameter of \qty{0.5}{mm} was pulled under laser heating into a sharp tip with an apex diameter of $\sim\qty{160}{nm}$. 
\qty{10}{nm} titanium/\qty{50}{nm} gold side contacts were deposited via electron beam evaporation, followed by an additional \qty{10}{nm} titanium/\qty{15}{nm} platinum strip positioned at $\sim\qty{0.5}{mm}$ from the tip apex that forms a shunt resistor of $\sim\qty{3}{\ohm}$. 
The contact pads were subsequently coated with a thick layer of indium solder to ensure reliable electrical contact at cryogenic temperatures. 
Finally, indium was deposited onto the tip in a home-built thermal evaporator held at $T<\qty{20}{K}$: first at \ang{110} relative to the tip axis to cover the two opposing contacts (45--\qty{60}{nm}), and then head-on (40--\qty{50}{nm}). 
This sequence produced a uniform, low-grain-size superconducting film interposed with two Dayem-bridge-type Josephson junctions. 
The resulting sensor exhibited a magnetic field sensitivity of $\sim\qty{10}{nT.Hz^{-1/2}}$, corresponding to a flux sensitivity of $\sim\qty{200}{n\Phi_0.Hz^{-1/2}}$ where $\Phi_0=h/2e$ is the superconducting magnetic flux quantum (the signal was read out using a series SQUID array amplifier\cite{Huber_DC_2001} in a flux-locked loop). 

During measurements, the sample was biased in a quasi-constant-current configuration (i.e., under a constant bias voltage through a series resistor of 10--\qty{100}{\kilo\ohm}). 
Bias was applied symmetrically from two sides to provide virtual grounding in the center of the active region. 
True square wave current ($f=\qty{1477.7}{Hz}$) was employed to maintain constant power dissipation and thereby suppress parasitic thermal response of nSOT. 
Application of a few \unit{\uA} of current, however, became difficult when R2G turned highly resistive or even insulating under very low $|n_\mathrm{e}|$ and high $D$; this regime is accordingly excluded from the phase diagrams of Fig.~\ref{fig:fig3}. 

\begin{widetext}
\subsubsection*{Measurement conditions for presented data}
\begin{ruledtabular}
\begin{tabular}{S[table-number-alignment = center]|S[table-number-alignment = center]|S[table-number-alignment = center]}
 {nSOT bias field (mT)} & {nSOT transfer function (V/T)} & {nSOT scan height (nm)} \\
 \hline
 29.3 & {30--}30.8 & 135 \\
\end{tabular}
\end{ruledtabular}
\begin{ruledtabular}
\begin{tabular}{c|S[table-number-alignment = center]|S[table-number-alignment = center]|S[table-number-alignment = center]}
 Data & {Bias voltage (V)} & {Series resistor (\unit{\kilo\ohm})} & {$I_0$ (\unit{\uA})} \\
 \hline
 Figure~\ref{fig:fig2}a,b, Extended Data Fig.~\ref{fig:ED_B_J_stream_fig2}a--d & 0.5 & 100 & 4.4 \\
 Figure~\ref{fig:fig2}c,d, Extended Data Fig.~\ref{fig:ED_B_J_stream_fig2}e--h & 0.5 & 100 & 4.6 \\ 
 Figure~\ref{fig:fig2}e,f, Extended Data Fig.~\ref{fig:ED_B_J_stream_fig2}i--l & 0.5 & 100 & 4.5 \\
 Figure~\ref{fig:fig3}a,b,c,f & 0.1 & 10 & {3.2--}6.9 \\
 Figure~\ref{fig:fig4}a, Extended Data Fig.~\ref{fig:ED_linearity}a & 0.5 & 100 & 4.6 \\
 Extended Data Fig.~\ref{fig:ED_linearity}b & 1.0 & 100 & 11 \\
 Figure~\ref{fig:fig4}b & 5 & 100 & 46 \\
 Figure~\ref{fig:fig4}c & 10 & 100 & 93 \\
 Figure~\ref{fig:fig4}g & 0.5 & 100 & 4.6 \\
 Figure~\ref{fig:fig4}h & 0.52 & 10 & 28 \\
 Figure~\ref{fig:fig4}i & 2.0 & 10 & 100 \\
\end{tabular}
\end{ruledtabular}
\end{widetext}

\subsection*{Reconstruction of current density and streamlines}

\paragraph*{Overview.}
Two-dimensional (2d) current density components $J_x(x,y)$ and $J_y(x,y)$, as well as the associated streamlines, were reconstructed from the measured fringe magnetic field $B(x,y)$ using a Fourier-space inversion algorithm\cite{Feldmann_Resolution_2004,Meltzer_Direct_2017}. 
The streamlines correspond to contours of the stream function $g(x,y)$, defined such that 
\begin{align}
    \nabla\times\left(g(x,y)\hat{z}\right)=\mathbf{J}(x,y).
    \label{eq:g2J}
\end{align}
Quantitatively, a net current of $\Delta I$ flows in the space confined by the contours $g(x,y)=I$ and $g(x,y)=I+\Delta I$. 
The scalar field $g(x,y)$ and the vector $\mathbf{J}(x,y)$ therefore contain equivalent information (up to an additive constant) and can be independently inverted from the measured $B(x,y)$. 

\paragraph*{Forward model.}
In Fourier space, the forward kernel propagating $g$ to the measured $B$ field is
\begin{widetext}
\begin{align}
    \hat{K_g}(k_x,k_y)=\frac{\mu_0}{2}\left(\sqrt{k_x^2+k_y^2}-ik_x\tan\theta\cos\varphi-ik_y\tan\theta\sin\varphi\right)e^{-\sqrt{k_x^2+k_y^2}h},
    \label{eq:g2B}
\end{align}
\end{widetext}
while the kernels connecting $\mathbf{J}$ to $B$ are 
\begin{subequations}
    \label{eq:J2B}
    \begin{align}
        \hat{K_x}(k_x,k_y)&=-i\frac{k_y}{k_x^2+k_y^2}\hat{K_g}(k_x,k_y),\\
        \hat{K_y}(k_x,k_y)&=i\frac{k_x}{k_x^2+k_y^2}\hat{K_g}(k_x,k_y).
    \end{align}
\end{subequations}
Here $h$ is the nSOT scan height and $\theta$ are $\varphi$ are tilt angles of the nSOT such that $B=B_z+\tan\theta\,(B_x\cos\varphi+B_y\sin\varphi)$. 
In our experiment a small unintentional tilt was present; $\theta=\ang{7.5}$ and $\varphi=\ang{0}$ provided good agreement between measured and simulated signals. 
The choice of $\hat{K_x}(k_x,k_y)$ and $\hat{K_y}(k_x,k_y)$ is not unique as current densities $J_x(x,y)$ and $J_y(x,y)$ are related by continuity in a steady state. 
Our formulation in \eqref{eq:J2B} has the benefit of ensuring this continuity in the inverse problem, as will be seen shortly. 
To account for finite nSOT size, all kernels were further multiplied by the point spread function 
\begin{align}
    \hat{f}(k_x,k_y)=\frac{2\mathcal{J}_1\left(\sqrt{k_x^2+k_y^2}r\right)}{\sqrt{k_x^2+k_y^2}r},
    \label{eq:PSF}
\end{align}
where $\mathcal{J}_1$ is the first-order Bessel function and $r$ is the effective nSOT radius. 

\paragraph*{Inverse problem and regularization.}
The quantities $\hat{g}(k_x,k_y)$, $\hat{J_x}(k_x,k_y)$ and $\hat{J_y}(k_x,k_y)$ in Fourier space were obtained through Tikhonov regularization with a Laplacian operator, 
\begin{align}
    \hat{g}(k_x,k_y)=\frac{\hat{K_g}^\ast(k_x,k_y)\hat{B}(k_x,k_y)}{\left|\hat{K_g}(k_x,k_y)\right|^2+\lambda_g(k_x^2+k_y^2)^2},
    \label{eq:B2g}
\end{align}
and 
\begin{widetext}
\begin{subequations}
    \label{eq:B2gJ}
    \begin{align}
        \hat{J_x}(k_x,k_y) &= \frac{\hat{K_x}^\ast(k_x,k_y)\hat{B}(k_x,k_y)}{\left|\hat{K_x}(k_x,k_y)\right|^2+\left|\hat{K_y}(k_x,k_y)\right|^2+\lambda_J(k_x^2+k_y^2)^2},\\
        \hat{J_y}(k_x,k_y) &= \frac{\hat{K_y}^\ast(k_x,k_y)\hat{B}(k_x,k_y)}{\left|\hat{K_x}(k_x,k_y)\right|^2+\left|\hat{K_y}(k_x,k_y)\right|^2+\lambda_J(k_x^2+k_y^2)^2}.
    \end{align}
\end{subequations}
\end{widetext}
Here $\lambda_g$ and $\lambda_J$ are regularization parameters that control the tradeoff between noise amplification (for small $\lambda$) and oversmoothing (for large $\lambda$). 
Optimal values for our datasets are $\lambda_g=\qty{5e-5}{T^2.\um^4.A^{-2}}$ and $\lambda_J=\qty{5e-6}{T^2.\um^6.A^{-2}}$. 
We note that the solutions in \eqref{eq:B2gJ} automatically satisfy current continuity, $ik_x\hat{J_x}(k_x,k_y)+ik_y\hat{J_y}(k_x,k_y)=0$. 

\paragraph*{Padding and background correction.}
Accurate Fourier-space inversion requires $B(x,y)$ to be defined over the full 2d plane. 
Because experimental data are finite in extent, we padded the measured $B(x,y)$ beyond the scan window under specific boundary assumptions. 
Two different schemes, \emph{Replication}---extending the edge values outward, and \emph{Reflection}---mirroring the data across the boundary, were found to produce less than \qty{3}{\percent} difference in normalized $\mathbf{J}(x,y)$ and $g(x,y)$. 

A further correction addresses the unconstrained $\mathbf{k}=0$ component of the inversion (where the kernels either vanish or are ill-defined), which can otherwise yield unphysical net currents across the scan region. 
We subtracted the mean value of $J_y(x,y)$ over the area outside the device boundary but within the measurement window, ensuring that the average background current vanishes. 
In some cases, we additionally subtracted a parabolic background fitted to $J_y$ to remove residual slow-varying artifacts. 
The stream function $g(x,y)$ was then shifted accordingly to maintain Eq.~\eqref{eq:g2J}. 

\subsubsection*{Simplified 1d reconstruction}
The above procedures yield full 2d maps of $\mathbf{J}(x,y)$ and $g(x,y)$ but require complete spatial maps of $B(x,y)$ as input. 
For flow concentration and counterflow plots in Fig.~\ref{fig:fig3}, we instead adopted a more efficient approach where $J_y(x,\pm y_0)$ and $I_x(x)$ are derived from magnetic field data along a one-dimensional (1d) trajectory. 
$J_y(x,\pm y_0)$ can be directly reconstructed from $B(x,\pm y_0)$, since these regions are far away from the lateral chambers and therefore translationally invariant in $y$ (i.e., $k_y=0$); here, the inversion reduces to 
\begin{align}
    \hat{J_y}(k_x)=\frac{\hat{K_y}^\ast(k_x,0)\hat{B}(k_x)}{\left|\hat{K_x}(k_x,0)\right|^2+\left|\hat{K_y}(k_x,0)\right|^2+\lambda_Jk_x^4}.
    \label{eq:B2Jy_1d}
\end{align}

Estimating $I(x)$ is more subtle. This quantity can be rewritten as 
\begin{align}
    I(x)\equiv\Delta g(x)=g(x,0)-\frac{g(x,-y_0)+g(x,y_0)}{2},
    \label{eq:IxDg}
\end{align}
which bears similarity to the corresponding field difference 
\begin{align}
    \Delta B(x)=B(x,0)-\frac{B(x,-y_0)+B(x,y_0)}{2}.
    \label{eq:DB}
\end{align}
In general, $\Delta g(x)$ and $\Delta B(x)$ are not uniquely connected, since the full kernel \eqref{eq:g2B} relating $g(x,y)$ to $B(x,y)$ depends on the transverse wavenumber $k_y$. 
However, the definition of $\Delta B(x)$, together with the spatial structure of $B(x,y)$, introduces a natural filter in $k_y$. 
Specifically, the effective window function suppresses long-wavelength components and peaks at finite $|k_y|$, thereby selecting a characteristic transverse wavenumber $|k_y|\approx 2\pi/l_y$ where $l_y$ denotes the $y$ scale of the chambers. 
Within this approximation, the inversion takes the form 
\begin{align}
    \hat{\tilde{\Delta g}}(k_x)=\frac{\hat{\tilde{K_g}}^\ast(k_x)\hat{\Delta B}(k_x)}{\left|\hat{\tilde{K_g}}(k_x)\right|^2+\lambda_gk_x^4},
    \label{eq:DB2Dg_1d}
\end{align}
with an effective 1d kernel 
\begin{widetext}
\begin{align}
    \hat{\tilde{K_g}}(k_x)=\frac{\mu_0}{2}\left(\sqrt{k_x^2+\left(\frac{2\pi}{l_y}\right)^2}-ik_x\tan\theta\cos\varphi\right)e^{-\sqrt{k_x^2+\left(\frac{2\pi}{l_y}\right)^2}h}.
    \label{eq:g2B_1d}
\end{align}
\end{widetext}
Here the $ik_y$ term in the original 2d kernel is dropped since the filtering preserves the $\pm k_y$ components symmetrically. 
Extended Data Figure~\ref{fig:ED_rec_2d1d} compares $\tilde{\Delta g}(x)$ reconstructed using this 1d approximation ($l_y=\qty{3}{\um}$) to the full 2d result, showing good agreement and validating the simplified procedure. 

\subsection*{Numerical simulations}
As a minimal model describing the device across the ballistic, hydrodynamic, and diffusive regimes, we consider a toy kinetic theory of one species of fermions in 2d with isotropic dispersion relation $\epsilon(|\mathbf{k}|)$, with a circular Fermi surface corresponding to $\epsilon_\mathrm{F}=\mu_0=\epsilon(|\mathbf k|)$. 
We work in linear response and assume $T\ll T_\mathrm{F}$, so that the single particle distribution function $f(\mathbf x,\mathbf k)$ is approximately
\begin{align}
    f(\mathbf x,\mathbf k)=\Theta(\mu_0-\epsilon(\mathbf k))-\phi(\mathbf x,\mathbf k)\delta(\mu_0-\epsilon(\mathbf k)),
    \label{eq:linearResponse}
\end{align}
where $\phi(\mathbf x, \mathbf k)$ parametrizes the linear-response deviation from equilibrium of the distribution function, $\Theta$ is the Heaviside step function, and $\delta$ is Dirac's delta function. 
It is convenient to take the momentum dependence in polar coordinates, $k_x=k\cos\theta,k_y=k\sin\theta$. 
Given Eq.~\eqref{eq:linearResponse}, we may then treat $\phi$ as dependent only on $\theta$, as the delta function fixes $k=k_\mathrm{F}$ [with $\epsilon_\mathrm{F}=\epsilon(k_\mathrm{F})$]. 
The Boltzmann transport equation for $f$ can be rewritten in terms of $\phi$ as
\begin{align}
    \partial_t\phi+v_\mathrm{F}\left(\cos\theta\,\partial_x+\sin\theta\,\partial_y\right)\phi = -W[\phi],
    \label{eq:linearBoltzmann}
\end{align}
where $v_\mathrm{F}$ is the Fermi velocity (in general, dependent on the electronic chemical potential $\mu_0$) and $W[\phi]$ is the collision integral describing the interaction of particles with their environment or each other. 
We have also assumed the absence of external forcing on the electrons. 

Assuming $\phi$ is small, we may linearize the equations around equilibrium $\phi=0$. In this regime, Eq.~\eqref{eq:linearBoltzmann} admits a natural Fourier decomposition 
\begin{widetext}
\begin{align}
    \phi=\frac{a_0(\mathbf{x},t)}{2}+\sum_{m=1}^\infty\left(a_m(\mathbf{x},t)\cos(m\theta))+b_m(\mathbf{x},t)\sin(m\theta)\right).
    \label{eq:fourierSeries}
\end{align}
For the collision matrix, we may use rotational symmetry to restrict the form of $W[\phi]$ as 
\begin{align}
    W[\phi]=v_\mathrm{F}\,l^{-1}_0\frac{a_0(\mathbf{x},t)}{2}+\sum_{m=1}^\infty v_\mathrm{F}\,l^{-1}_m\left(a_m(\mathbf{x},t)\cos(m\theta))+b_m(\mathbf{x},t)\sin(m\theta)\right),
\end{align}
where $l_n$ represents a mean free path for the $n^{\mathrm{th}}$ Fourier harmonic. 
In steady state, Eq.~\eqref{eq:linearBoltzmann} becomes an infinite set of coupled partial differential equations  
\begin{subequations}
    \label{eq:harmonicEquations}
    \begin{align}
        \partial_xa_1+\partial_yb_1 &= -l^{-1}_0a_0, \\
        \frac{1}{2}\left[\partial_x(a_{n+1}+a_{n-1})+\partial_y(b_{n+1}-b_{n-1})\right] &= -l^{-1}_na_n,\ n\geq1, \\
        \frac{1}{2}\left[\partial_x(b_{n+1}+b_{n-1})-\partial_y(a_{n+1}-a_{n-1})\right] &= -l^{-1}_nb_n,\ n\geq1.
    \end{align}
\end{subequations}
\end{widetext}
Here $b_0\equiv0$. 
We further adopt a two-time collision integral with $\ell_\mathrm{cr}$ representing the mean free path between current-relaxing collisions and $\ell_\mathrm{cc}$ that between current-conserving collisions, namely 
\begin{subequations}
    \label{eq:relaxationRates}
    \begin{align}
        l^{-1}_0 &= 0, \\
        l^{-1}_1 &= \ell^{-1}_\mathrm{cr}, \\
        l^{-1}_{n} &= \ell^{-1}_\mathrm{cr}+\ell^{-1}_\mathrm{cc},\ n\geq2.
    \end{align}
\end{subequations}
This corresponds to a linearized BGK model\cite{bhatnagarModelCollisionProcesses1954, mendlDyakonovShurInstabilityBallistictohydrodynamic2018}. 
We note that considering separate $l_n$ for even and odd $n$ would be straightforward and capture tomographic physics expected to be relevant for 2d electron liquids\cite{ledwithTomographicDynamicsScaleDependent2019}, but we find that two fit parameters are already sufficient to quantitatively model the present experiment. 

We solve Eqs.~\eqref{eq:harmonicEquations} numerically\cite{farrell2026characterizingelectronicscatteringrates,farrell2026FermiSea} using discontinuous Galerkin spectral element methods (DGSEM) as implemented in the \texttt{Trixi.jl} package for the \texttt{Julia} programming language\cite{ranocha2022adaptive,schlottkelakemper2021purely}. 
In particular, the solution is spatially reconstructed as a degree $4$ polynomial within each cell of an unstructured quadrilateral mesh with average grid size $\mathrm{d}x=\mathrm{d}y=\qty{100}{nm}$. 
The high-order reconstruction allows us to obtain accurate numerical solutions which capture the complex flows in the experimental geometry, despite the relatively coarse mesh. 
The Fourier series \eqref{eq:fourierSeries} are truncated at $n=n_{\max}$, and we have run simulations with a high $n_{\max}=50$ to quantitatively capture ballistic physics. 
The surface flux is local Lax--Friedrichs. 
To find the stationary solution, we integrate forward Eq.~\eqref{eq:linearBoltzmann} in time using an explicit Carpenter--Kennedy Runge--Kutta scheme until convergence to a time-independent solution is reached. 
Convergence is determined by the maximum residual, i.e., the $\infty$-norm of the time derivative $\left\lVert\frac{\mathrm{d}\phi}{\mathrm{d}t}\right\rVert_\infty\equiv\max\left|\frac{\mathrm{d}\phi}{\mathrm{d}t}\right|$. 

We have considered two types of boundary condition on the wall of the device, namely diffuse scattering and specular reflection. 
In either case, we must constrain only degrees of freedom (particles) that are \emph{incoming} to the domain. 
In the diffuse case, we choose $a_n$ and $b_n$ at the walls such that (at appropriately discretized angles) the value of $\phi(\theta)$ is uniform at every angle $\theta$ corresponding to a particle propagating away from the boundary. 
This constant value is chosen to ensure that no particles are emitted or absorbed by the boundary. 
In the case of specular reflection, the incoming degrees of freedom are instead constrained to satisfy $\phi(\theta)=\phi(\pi-\theta)$, which constrains the distribution at an outgoing angle in terms of an ingoing angle. 

On the source and drain contacts, we also constrain only the angles corresponding to particles entering the domain from the boundary. 
In this case, we again pick $a_n$ and $b_n$ such that the value of $\phi(\theta)$ is uniform at all angles corresponding to particles entering the domain, with constant value this time held \emph{fixed} rather than chosen to guarantee zero particles injected or absorbed at the boundary. 
This choice models an ideal (absorbing) Ohmic contact held at a fixed potential. 

\subsection*{Least squares fitting}

Our Boltzmann model takes $(\ell_\mathrm{cr},\ell_\mathrm{cc})$ and the sample geometry as input, and generates $\mathbf{J}(x,y)$ maps that are exact within model assumptions. 
Fitting to magnetometry data was performed \emph{in the magnetic field space} to eliminate finite resolution systematics (finite scan height, finite nSOT size, regularization, etc.) and Fourier-domain artifacts (ringing) associated with experimentally derived current maps. 
In other words, we computed theoretical $B(x,y)$ from simulated $\mathbf{J}(x,y)$ using the forward kernels \eqref{eq:J2B} (i.e., Biot--Savart law), which we then compared to the measured $B$ signal. 

According to Eqs.~\eqref{eq:B2Jy_1d} and \eqref{eq:DB2Dg_1d}, the reconstructed $J_y(x,\pm y_0)$ and $I_x(x)$ curves are related to $B(x,\pm y_0)$ and $\Delta B(x)$, respectively. 
We therefore define two per-pixel root mean square deviations (RMSDs): $\mathrm{RMSD}(J_y)$, computed from $B(x,\pm y_0)$ over the range $\qty{-1.2}{\um}\leq x\leq\qty{1.2}{\um}$, and $\mathrm{RMSD}(I_x)$, computed from $\Delta B(x)$ over $\qty{-1.8}{\um}\leq x\leq\qty{1.8}{\um}$. 
In both cases, the field curve is normalized by $B_\mathrm{range}\equiv B_{\max}-B_{\min}$ of the scan frame. 
The best-fit parameters were obtained by minimizing the (equally weighted) combined RMSD~$:=\sqrt{\mathrm{RMSD}(J_y)^2+\mathrm{RMSD}(I_x)^2}$. 

We find that $\mathrm{RMSD}(J_y)$ and $\mathrm{RMSD}(I_x)$ predominantly probe channel flow and vortex profiles, respectively, and that their combination is necessary in determining the exact transport regime. 
At the CNP, for example, $\mathrm{RMSD}(J_y)$ is small over a broad range of $\ell_\mathrm{cr}$ and $\ell_\mathrm{cc}$ (Extended Data Fig.~\ref{fig:ED_fit_Jy_Ix}a); both diffusive (green) and ballistic (purple) regimes produce a flat current profile, consistent with the experimental data (Extended Data Fig.~\ref{fig:ED_fit_Jy_Ix}b). 
A good fit for $\Delta B(x)$ (and thus $I_x(x)$), however, is achieved only in the diffusive regime and not in the ballistic regime (Extended Data Fig.~\ref{fig:ED_fit_Jy_Ix}c,d). 
At high density, in contrast, $\mathrm{RMSD}(I_x)$ favors instead the ballistic regime, consistent with the observed strong vortical flow (Extended Data Fig.~\ref{fig:ED_fit_Jy_Ix}g,h). 
Finally in the low-$|n_\mathrm{e}|$, high-$D$ flat band regime, $\mathrm{RMSD}(I_x)$ allows two scenarios (Extended Data Fig.~\ref{fig:ED_fit_Jy_Ix}k,l): the hydrodynamic regime (orange), or the direct crossover between diffusive and ballistic regimes in which $\ell_\mathrm{cc}$ remains large (blue). 
Here, $B(x)$ ($J_y(x)$) is crucial in the identification of hydrodynamic transport which alone can generate the concentrated Poiseuille flow (Extended Data Fig.~\ref{fig:ED_fit_Jy_Ix}i,j). 

\subsubsection*{Origins of unphysical current artifacts}
The experimental current maps of Fig.~2a--f exhibit an apparent finite current outside the device boundary, which cannot by physical. 
To understand the origin of this artifact, we subject the theoretical current maps to the same analysis chain applied to the data; as we show, the unphysical currents are an expected result of our finite resolution and finite signal-to-noise ratio which are most prominent when the current density is very low. 
Extended Data Figure~\ref{fig:ED_sim_vs_rec}a--d show microscopic current distributions in the diffusive, ballistic, and hydrodynamic regimes calculated from the linearized Boltzmann model. 
The expected magnetic field signal generated from these microscopic currents is shown in Extended Data Fig.~\ref{fig:ED_sim_vs_rec}e--g; its spatial derivatives are shown in Extended Data Fig.~\ref{fig:ED_sim_vs_rec}h--m. 
For these simulations, we account for both the finite height of the nSOT sensor and its finite diameter. 
Notably, neither $B(x,y)$ nor $\mathrm{d}B(x,y)/\mathrm{d}x$ and $\mathrm{d}B(x,y)/\mathrm{d}y$ vanishes outside the device area, a feature of magnetostatics consistent with experimental maps in Extended Data Fig.~\ref{fig:ED_B_J_stream_fig2}. 
We then infer the current maps \emph{from the theoretical $B$ fields} by applying the identical Fourier-domain reconstruction procedure as used for the experimental data (Extended Data Fig.~\ref{fig:ED_sim_vs_rec}n--q). 
The key distinguishing features of the three transport regimes are preserved by this analysis, including current concentration and vortical flow. 
However, unphysical currents are inevitably generated outside the device area due to finite resolution of the probe as well as Fourier-domain numerical artifacts. 
We note that it is important to compare data and models of the same quantity: our fitting is thus performed at the level of the fringe magnetic field itself, while the comparisons between theory and experiment of $J_y(x)/J_0$ and $I_x(x)/I_0$ shown in Extended Data Fig.~\ref{fig:ED_fit_Jy_Ix} are both the reconstructed, rather than microscopic physical currents. 

Noise also plays a more deleterious role where signals are small. 
To simulate its effects, we add a small amount of noise (\qty{30}{nT} per $\qty{25}{nm}\times\qty{25}{nm}$ pixel) to the fringe field simulated for total current $I_0=\qty{4.5}{\uA}$. 
Extended Data Figure~\ref{fig:ED_sim_vs_rec}r--u show the inferred currents in this case. 
Evidently, the presence of noise generates additional artifacts that closely resemble unphysical features in the experimental current maps, though again leaving the key features intact. 
We conclude that unphysical streamlines are indeed expected and inevitable artifacts of the finite probe resolution and finite signal-to-noise ratio. 

\subsection*{Effective mass dependence of $\ell_\mathrm{ee}$}

Here we provide a derivation of Eq.~\eqref{eq1}. 
According to the Fermi liquid theory\cite{Hui_Hydrodynamics_2025}, 
\begin{align}
    \ell_\mathrm{ee}=v_\mathrm{F}\tau_\mathrm{ee}\sim\frac{\hbar v_\mathrm{F}E_\mathrm{F}}{(k_\mathrm{B}T)^2},
    \label{eq:lee_FL}
\end{align}
where $v_\mathrm{F}$ is the Fermi velocity and $E_\mathrm{F}$ is the Fermi energy. 
In a parabolic band, 
\begin{subequations}
\label{eq:vF_EF}
\begin{align}
    v_\mathrm{F}&=\frac{\hbar k_\mathrm{F}}{m^*},\\
    E_\mathrm{F}&=\frac{\hbar^2k_\mathrm{F}^2}{2m^*},
\end{align}
\end{subequations}
where $k_\mathrm{F}=\sqrt{4\pi|n_\mathrm{e}|/g}$ is the Fermi wavelength and $g$ is the degeneracy ($g=4$ for R2G). 
Combining \eqref{eq:lee_FL},~\eqref{eq:vF_EF} yields 
\begin{align}
    \ell_\mathrm{ee}\sim\frac{\hbar^4}{2{m^*}^2(k_\mathrm{B}T)^2}\left(\frac{4\pi|n_\mathrm{e}|}{g}\right)^{3/2};
    \label{eq:lee_full}
\end{align}
dropping the prefactors gives us \eqref{eq1}. 

We note that this derivation is quantitatively valid only for a Galilean invariant single band system---which dual-gated R2G is not---but we still find qualitative, order-of-magnitude agreement between Eq.~\ref{eq:lee_full} and our data. 
For the flow profile presented in Fig.~\ref{fig:fig2}e,f, we have $m^*\approx0.5m_\mathrm{e}$ from the tight-binding model and $\ell_\mathrm{ee}\approx\ell_\mathrm{cc}\approx\qty{30}{nm}$ from the Boltzmann fit; plugging those into \eqref{eq:lee_full} yields $T\sim\qty{5}{K}$, broadly consistent with the cryostat temperature. 

\subsection*{Quantifying vortex deformation}

To quantify the current-driven deformation of vortical flow, we analyzed vortex center position(s), $y_\mathrm{c}$, marked by local extrema of the current stream function $g(x,y)$ (maxima for counterclockwise vortices and minima for clockwise vortices). 
Figure~\ref{fig:fig4}k shows the evolution of $y_\mathrm{c}$ for both chambers upon increasing $I_0$; error bars are defined by $|g-g_0|<0.001I_0$ where $g_0$ is the extremum value. 
Small values of $I_0$ give $y_\mathrm{c}\approx0$, i.e., vortices centered near the central horizontal axis of the device. 
Under large $I_0$, in contrast, the vortical flow assumes a crescent shape with two separate centers at the top and bottom halves of the scan frame (see, for example, Fig.~\ref{fig:fig4}i). 

A similar analysis of vortex center position(s) was performed on linear-Boltzmann theoretical flow patterns (as inferred from magnetic inversion) and presented in Extended Data Fig.~\ref{fig:ED_center}. 
We find $|y_\mathrm{c}|<\qty{40}{nm}$ in the ballistic and hydrodynamic regimes, i.e., vortices are always on the central axis; this implies the observation under high current drive goes beyond the range of validity of the linear Boltzmann model. 
We note, however, that the algorithm for determining $y_\mathrm{c}$ could find spurious local extrema in the diffusive regime where no vortices are expected; they are masked out (white region in Extended Data Fig.~\ref{fig:ED_center}b) by requiring that the error bars of $y_\mathrm{c}$ do not exceed the chamber size. 

\section*{Acknowledgments} 
The authors acknowledge discussions with L. Levitov, L. I. Glazman, B. Skinner, L. Balents, E. Berg, K. G. Nazaryan, and A. Panigrahi. 
Experimental work at UCSB was primarily supported by the Department of Energy under Award DE-SC0020043 to A.F.Y. 
Additional funding for nSOT sensor and microscope development was provided by the Experimental Physics Investigator program of the Gordon and Betty Moore Foundation under Award GMBF-13801 to A.F.Y. 
S.K. and A.C.B.J. acknowledge support from the Gordon and Betty Moore Foundation’s EPiQS Initiative via Grant GBMF-10279. 
K.W. and T.T. acknowledge support from the JSPS KAKENHI (Grant Numbers 21H05233 and 23H02052), the CREST (JPMJCR24A5), JST and World Premier International Research Center Initiative (WPI), MEXT, Japan. 
Theoretical modeling at CU Boulder was supported by the National Science Foundation under CAREER Grant DMR-2145544. 
This work made use of shared equipment supported by the National Science Foundation through Enabling Quantum Leap: Convergent Accelerated Discovery Foundries for Quantum Materials Science, Engineering and Information (QAMASE-i) Award DMR-1906325. 

\section*{Author contributions}
A.F.Y. conceived and supervised the project. 
H.S., S.K., L.H. and Y.C. designed and fabricated the device. 
T.T. and K.W. grew the hBN crystals. 
D.G. and A.K. prepared nanoSQUID-on-tip sensors. 
C.Z. and E.R. performed the measurements and data analysis. 
C.Z. optimized the current reconstruction algorithm. 
J.H.F. developed the numerical methods. 
J.H.F. and A.L. provided theoretical support. 
M.E.H. provided SQUID array amplifiers for nSOT readout. 
C.Z., E.R. and A.F.Y. wrote the paper with inputs from all other authors. 

\section*{Competing interests}
The authors declare no competing interests.

\section*{Data and code availability}
All supporting data and numerical codes for this paper and other findings of this study are available from the corresponding author upon request. 

% Extended Data Figures
\clearpage
\setcounter{figure}{0}
\renewcommand{\figurename}{\textbf{Extended Data Fig.}}

\begin{figure*}[ht!]
\centering
\includegraphics[width=5in]{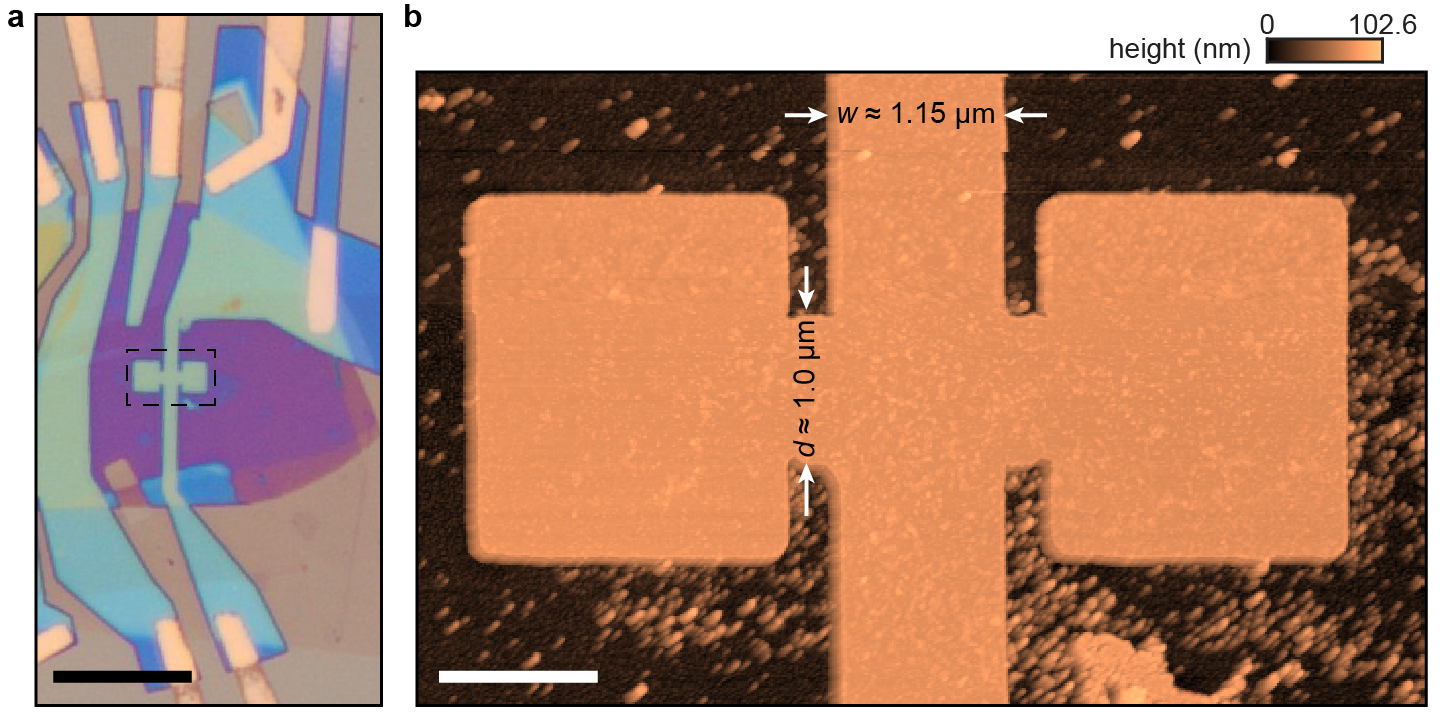}
\caption{\textbf{Sample geometry.}
\textbf{a}, Optical image of dual-gated R2G after patterning. Scale bar: \qty{10}{\um}. 
\textbf{b}, Zoomed-in atomic force microscopy (AFM) image of the active region. Scale bar: \qty{1}{\um}. 
}
\label{fig:ED_sample}
\end{figure*}

\begin{figure*}[ht!]
\centering
\includegraphics[width=7in]{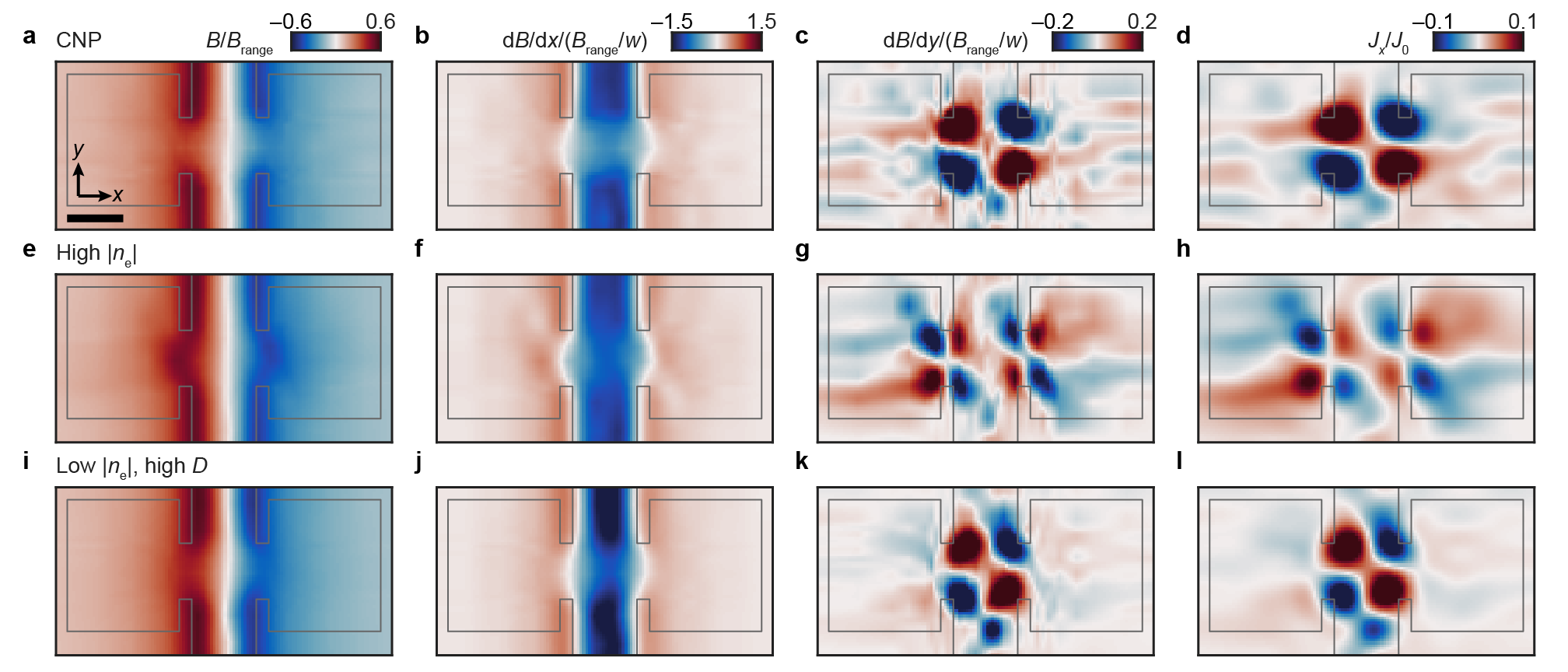}
\caption{\textbf{Magnetic field, field derivatives, and current density for the data of Fig.~\ref{fig:fig2}.}
\textbf{a}, Fringe magnetic field signal $B(x,y)$ measured at $n_\mathrm{e}=\qty{0e12}{cm^{-2}}$ and $D=\qty{0}{V/nm}$. Scale bar: \qty{1}{\um}. $B(x,y)$ is normalized by $B_\mathrm{range}\equiv B_{\max}-B_{\min}$. 
\textbf{b}, $\mathrm{d}B/\mathrm{d}x$ obtained from the data of panel a and normalized by $B_\mathrm{range}/w$. 
\textbf{c}, $\mathrm{d}B/\mathrm{d}y$ obtained from the data of panel a. 
\textbf{d}, Reconstructed current density $J_x(x,y)$ from the data of panel a. $J_x$ is normalized by $J_0\equiv I_0/w$. 
\textbf{e}, $B(x,y)$ at $n_\mathrm{e}=\qty{-1.3e12}{cm^{-2}}$ and $D=\qty{0.39}{V/nm}$. 
\textbf{f}, $\mathrm{d}B/\mathrm{d}x$ obtained from the data of panel e. 
\textbf{g}, $\mathrm{d}B/\mathrm{d}y$ obtained from the data of panel e. 
\textbf{h}, $J_x(x,y)$ reconstructed from the data of panel e. 
\textbf{i}, $B(x,y)$ at $n_\mathrm{e}=\qty{-0.4e12}{cm^{-2}}$ and $D=\qty{0.39}{V/nm}$. 
\textbf{j}, $\mathrm{d}B/\mathrm{d}x$ obtained from the data of panel i. 
\textbf{k}, $\mathrm{d}B/\mathrm{d}y$ obtained from the data of panel i. 
\textbf{l}, $J_x(x,y)$ reconstructed from the data of panel i. 
}
\label{fig:ED_B_J_stream_fig2}
\end{figure*}

\begin{figure*}[ht!]
\centering
\includegraphics[width=7in]{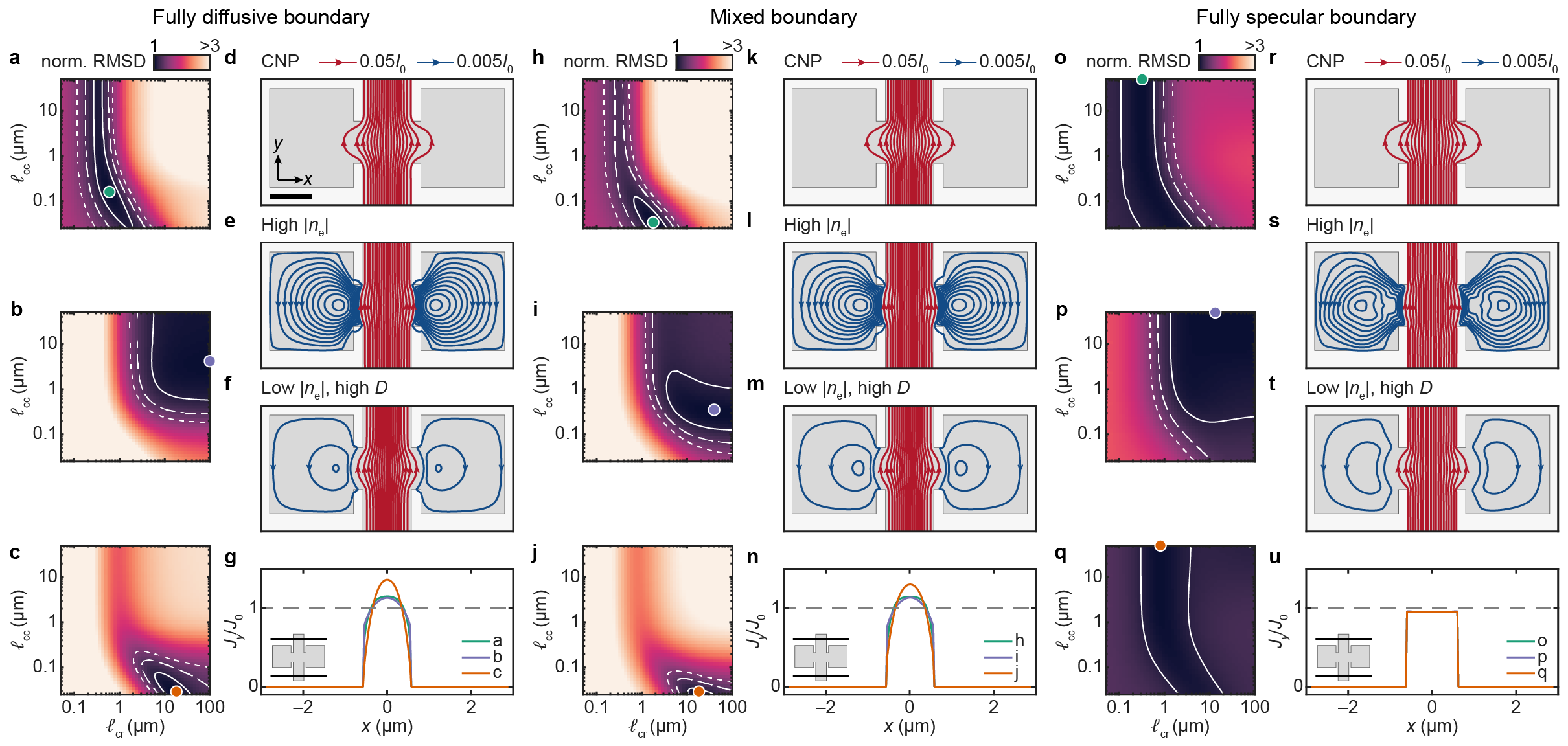}
\caption{\textbf{Boltzmann fitting under different boundary conditions.}
\textbf{a}, Combined RMSD between experimental data at $n_\mathrm{e}=\qty{0e12}{cm^{-2}}$ and $D=\qty{0}{V/nm}$ and the linearized Boltzmann model parameterized by $\ell_\mathrm{cr}$ and $\ell_\mathrm{cc}$ with a fully diffusive boundary. The RMSD is normalized by the minimum value, indicated by the dot and corresponding to the best fit. Contours of 1.1, 1.3, and 1.5 delineate phenomenological thresholds. 
\textbf{b}, RMSD for $n_\mathrm{e}=\qty{-1.3e12}{cm^{-2}}$ and $D=\qty{0.39}{V/nm}$ with a diffusive boundary. 
\textbf{c}, RMSD for $n_\mathrm{e}=\qty{-0.4e12}{cm^{-2}}$ and $D=\qty{0.39}{V/nm}$ with a diffusive boundary.
\textbf{d}, Theoretical current streamlines for $\ell_\mathrm{cr}=\qty{600}{nm}$ and $\ell_\mathrm{cc}=\qty{161}{nm}$, corresponding to the best fit in panel a. Scale bar: \qty{1}{\um}. 
\textbf{e}, Theoretical current streamlines for $\ell_\mathrm{cr}=\qty{100}{\um}$ and $\ell_\mathrm{cc}=\qty{4.2}{\um}$, corresponding to the best fit in panel b. 
\textbf{f}, Theoretical current streamlines for $\ell_\mathrm{cr}=\qty{18.2}{\um}$ and $\ell_\mathrm{cc}=\qty{29}{nm}$, corresponding to the best fit in panel c. 
\textbf{g}, $J_y(x)/J_0$ at $y=\pm\qty{2}{\um}$ for the best fits described in panels d, e, and f. 
Panels a--g reproduce Fig.~\ref{fig:fig2}i--o. 
\textbf{h}, RMSD for $n_\mathrm{e}=\qty{0e12}{cm^{-2}}$ and $D=\qty{0}{V/nm}$ with a half-diffusive, half-specular boundary. 
\textbf{i}, RMSD for $n_\mathrm{e}=\qty{-1.3e12}{cm^{-2}}$ and $D=\qty{0.39}{V/nm}$ with a mixed boundary. 
\textbf{j}, RMSD for $n_\mathrm{e}=\qty{-0.4e12}{cm^{-2}}$ and $D=\qty{0.39}{V/nm}$ with a mixed boundary. 
\textbf{k}, Theoretical current streamlines for $\ell_\mathrm{cr}=\qty{1.78}{\um}$ and $\ell_\mathrm{cc}=\qty{34}{nm}$, corresponding to the best fit in panel h. 
\textbf{l}, Theoretical current streamlines for $\ell_\mathrm{cr}=\qty{39}{\um}$ and $\ell_\mathrm{cc}=\qty{350}{nm}$, corresponding to the best fit in panel i. 
\textbf{m}, Theoretical current streamlines for $\ell_\mathrm{cr}=\qty{18.2}{\um}$ and $\ell_\mathrm{cc}=\qty{29}{nm}$, corresponding to the best fit in panel j. 
\textbf{n}, $J_y(x)/J_0$ at $y=\pm\qty{2}{\um}$ for the best fits described in panels k, l, and m. 
Compared to the diffusive boundary, the mixed boundary results in slightly lower values of $\ell_\mathrm{cc}$ but maintains the broad contrast between the three regimes. 
The diffusive boundary is adopted in the main text both for simplicity and because it leads to a more conservative estimation of $\ell_\mathrm{cc}$, thus strengthening the conclusion of an ultralow $\ell_\mathrm{cc}$ hydrodynamic regime. 
\textbf{o}, RMSD for $n_\mathrm{e}=\qty{0e12}{cm^{-2}}$ and $D=\qty{0}{V/nm}$ with a fully specular boundary. 
\textbf{p}, RMSD for $n_\mathrm{e}=\qty{-1.3e12}{cm^{-2}}$ and $D=\qty{0.39}{V/nm}$ with a specular boundary. 
\textbf{q}, RMSD for $n_\mathrm{e}=\qty{-0.4e12}{cm^{-2}}$ and $D=\qty{0.39}{V/nm}$ with a specular boundary. 
\textbf{r}, Theoretical current streamlines for $\ell_\mathrm{cr}=\qty{320}{nm}$ and $\ell_\mathrm{cc}=\qty{50}{\um}$, corresponding to the best fit in panel o. 
\textbf{s}, Theoretical current streamlines for $\ell_\mathrm{cr}=\qty{13.3}{\um}$ and $\ell_\mathrm{cc}=\qty{50}{\um}$, corresponding to the best fit in panel p. 
\textbf{t}, Theoretical current streamlines for $\ell_\mathrm{cr}=\qty{820}{nm}$ and $\ell_\mathrm{cc}=\qty{50}{\um}$, corresponding to the best fit in panel q. 
\textbf{u}, $J_y(x)/J_0$ at $y=\pm\qty{2}{\um}$ for the best fits described in panels r, s, and t. 
The specular boundary collapses distinction in the channel flow profile and in particular fails to reproduce the observed Poiseuille flow in the flat band regime, and is thus ruled out. 
}
\label{fig:ED_boundary}
\end{figure*}

\begin{figure*}[ht!]
\centering
\includegraphics[width=6in]{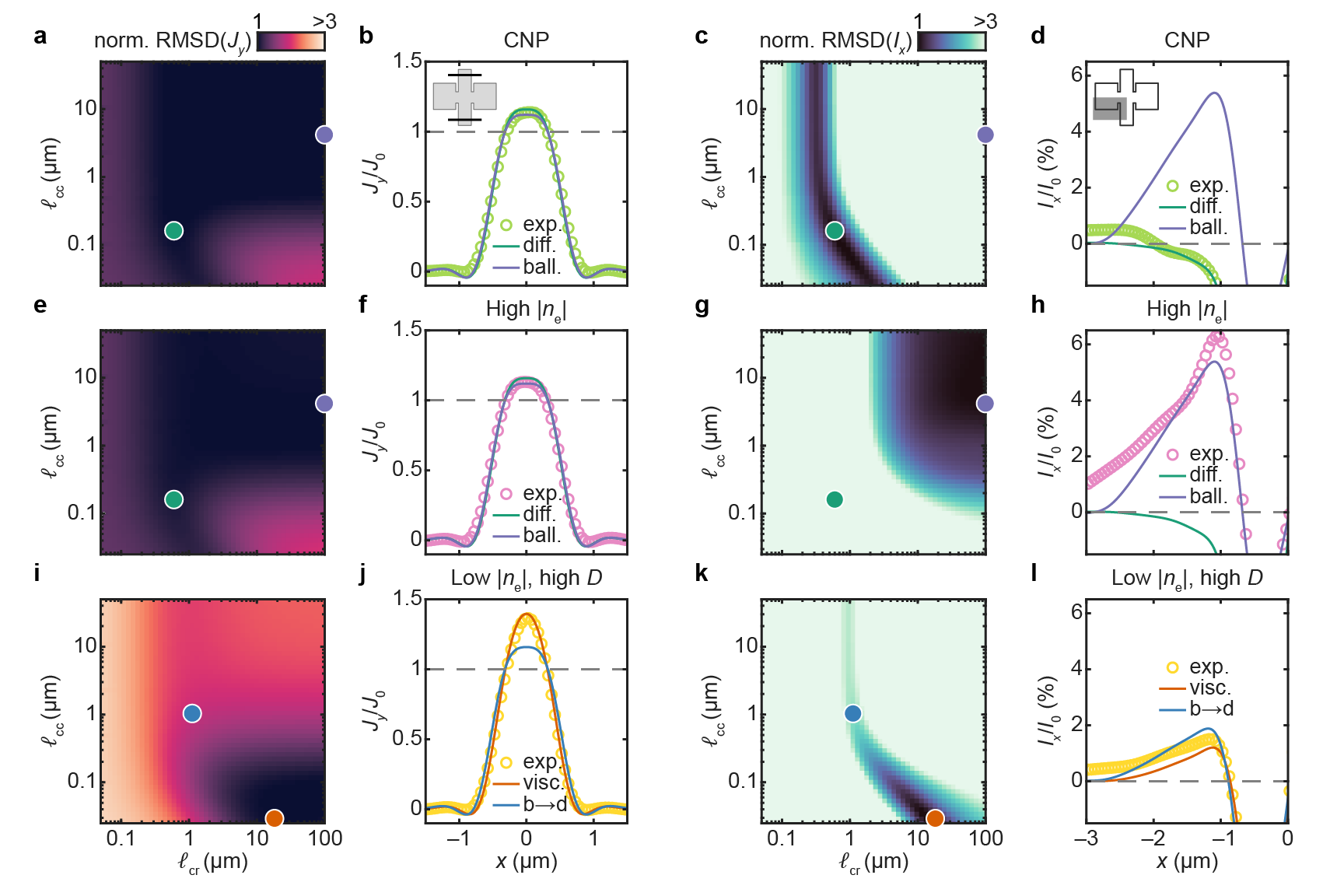}
\caption{\textbf{Least squares fitting of $J_y(x)$ and $I_x(x)$.}
\textbf{a}, Root mean square deviation (RMSD) for $J_y(x)/J_0$ between experimental data at $n_\mathrm{e}=\qty{0e12}{cm^{-2}}$ and $D=\qty{0}{V/nm}$ and the linearized Boltzmann model parameterized by $\ell_\mathrm{cr}$ and $\ell_\mathrm{cc}$. The RMSD is normalized by the minimum value. 
\textbf{b}, Comparison between experimental $J_y(x)/J_0$ and fits in the diffusive and ballistic regimes as inferred from magnetic inversion, using parameters corresponding to dots in panel a. 
\textbf{c}, Normalized RMSD for $I_x(x)/I_0$ at the same $n_\mathrm{e}$ and $D$ as panel a. 
\textbf{d}, Comparison between experimental $I_x(x)/I_0$ and fits in the diffusive and ballistic regimes. 
\textbf{e}, Normalized RMSD for $J_y(x)/I_0$ at $n_\mathrm{e}=\qty{-1.3e12}{cm^{-2}}$ and $D=\qty{0.39}{V/nm}$. 
\textbf{f}, Comparison between experimental $J_y(x)/J_0$ and fits in the diffusive and ballistic regimes. 
\textbf{g}, Normalized RMSD for $I_x(x)/I_0$ at the same $n_\mathrm{e}$ and $D$ as panel e. 
\textbf{h}, Comparison between experimental $I_x(x)/I_0$ and fits in the diffusive and ballistic regimes. 
\textbf{i}, Normalized RMSD for $J_y(x)/I_0$ at $n_\mathrm{e}=\qty{-0.4e12}{cm^{-2}}$ and $D=\qty{0.39}{V/nm}$. 
\textbf{j}, Comparison between experimental $J_y(x)/J_0$ and fits in the viscous (i.e., hydrodynamic) regime as well as at the direct transition between ballistic and diffusive regimes, using parameters corresponding to dots in panel i. 
\textbf{k}, Normalized RMSD for $I_x(x)/I_0$ at the same $n_\mathrm{e}$ and $D$ as panel i. 
\textbf{l}, Comparison between experimental $I_x(x)/I_0$ and fits in the viscous regime and at the ballistic-to-diffusive transition. 
}
\label{fig:ED_fit_Jy_Ix}
\end{figure*}

\begin{figure*}[ht!]
\centering
\includegraphics[width=7in]{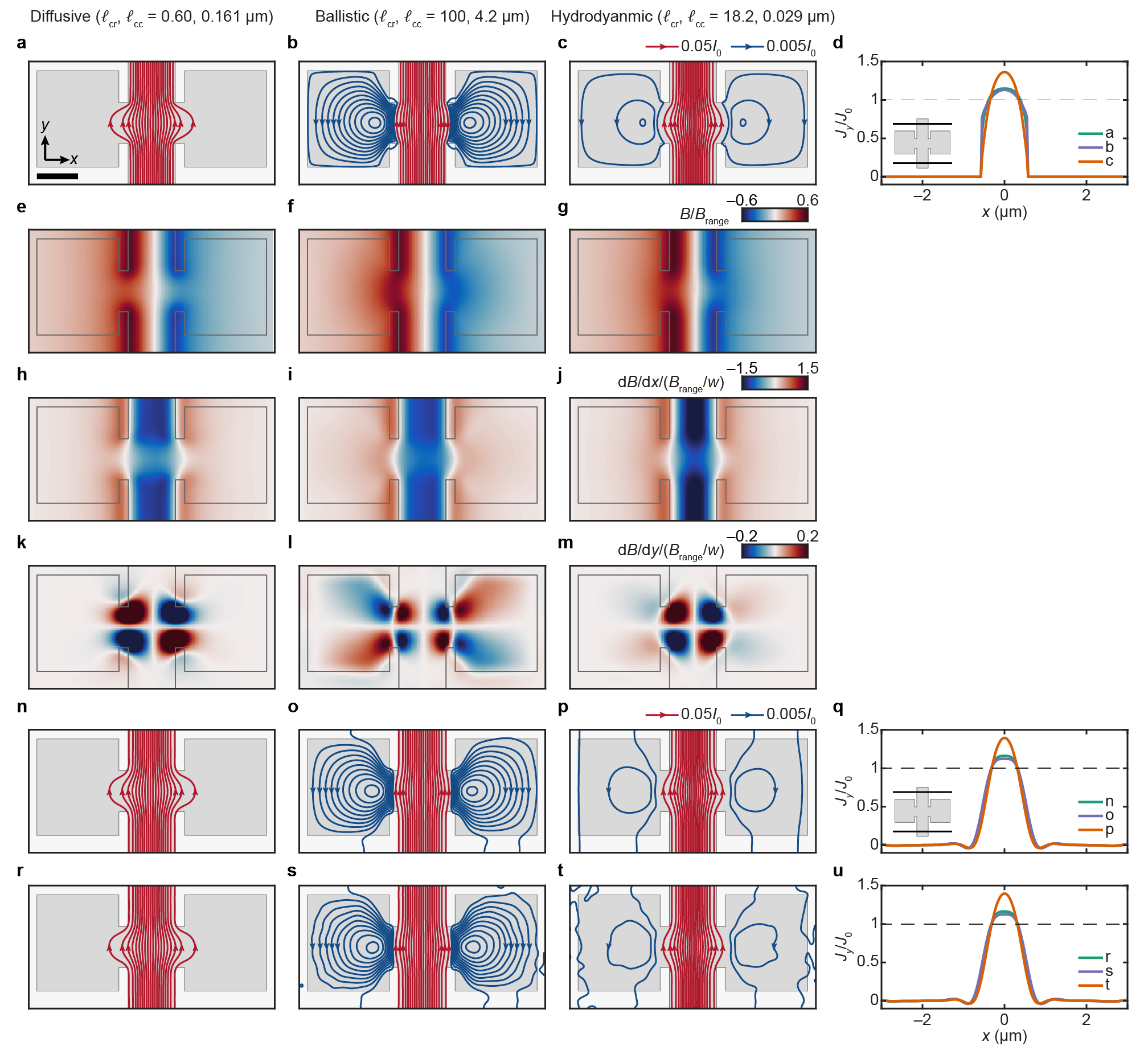}
\caption{\textbf{Simulating artifacts in reconstructed current maps.}
\textbf{a}, Physical current streamlines calculated from the Boltzmann model in the diffusive regime, for $\ell_\mathrm{cr}=\qty{600}{nm}$ and $\ell_\mathrm{cc}=\qty{161}{nm}$. Scale bar: \qty{1}{\um}. 
\textbf{b}, Streamlines in the ballistic regime, for $\ell_\mathrm{cr}=\qty{100}{\um}$ and $\ell_\mathrm{cc}=\qty{4.2}{\um}$. 
\textbf{c}, Streamlines in the hydrodynamic regime, for $\ell_\mathrm{cr}=\qty{18.2}{\um}$ and $\ell_\mathrm{cc}=\qty{29}{nm}$. 
\textbf{d}, Line traces of $J_y(x,\pm\qty{2}{\um})/J_0$ corresponding to panels a--c. 
\textbf{e}--\textbf{g}, Simulated $B(x,y)$ as measured by a \qty{160}{nm}-diamter nSOT sensor \qty{135}{nm} above the sample corresponding to the physical currents in panels a--c. 
\textbf{h}--\textbf{j}, $\mathrm{d}B(x,y)/\mathrm{d}x$ derived from the data in panels e--g. 
\textbf{k}--\textbf{m}, $\mathrm{d}B(x,y)/\mathrm{d}y$ derived from the data in panels e--g. 
\textbf{n}--\textbf{p}, Current streamlines as would be determined from the field maps in panels e--g using standard Fourier-domain methods. Note the unphysical streamlines outside the device area. 
\textbf{q}, $J_y(x,\pm\qty{2}{\um})/J_0$ for the data in panels n--p. 
\textbf{r}--\textbf{t}, Current streamlines as would be determined from the field maps in panels e--g using standard Fourier-domain methods, assuming an additional uncorrelated random noise of \qty{30}{nT} per $(\qty{25}{nm})^2$ pixel. Note the effect of the noise is largest where the streamlines are the least dense. 
\textbf{u}, $J_y(x,\pm\qty{2}{\um})/J_0$ for the data in panels r--t. 
}
\label{fig:ED_sim_vs_rec}
\end{figure*}

\begin{figure*}[ht!]
\centering
\includegraphics[width=6in]{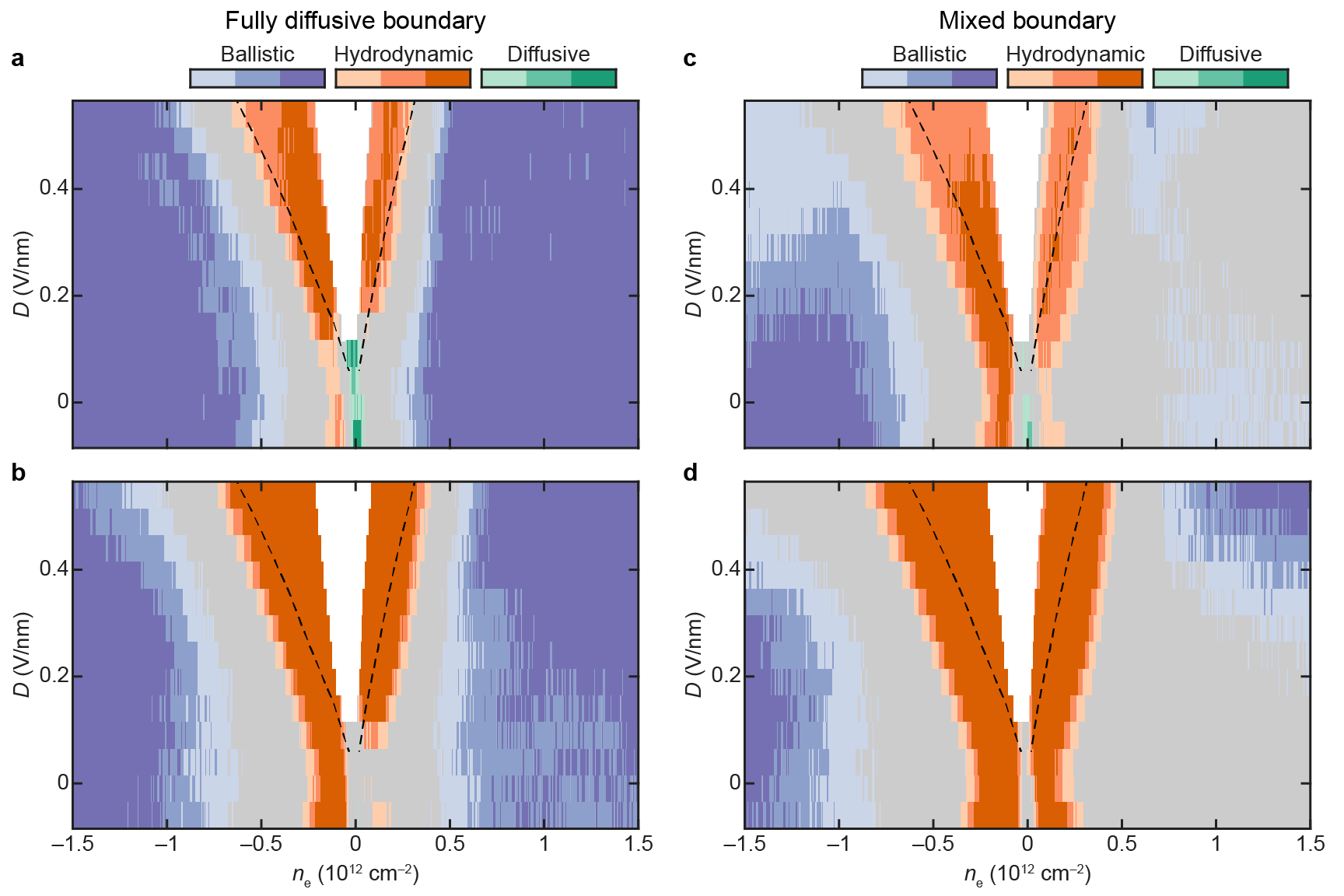}
\caption{\textbf{Boundary- and threshold-dependent phase diagram.}
\textbf{a}, Phase diagram as shown in Fig.~\ref{fig:fig3}i, obtained from fitting to the Boltzmann model with a fully diffusive boundary. Blue represents the ballistic regime where $\min{\ell_\mathrm{cr}}>\qty{2}{\um}$ and $\min{\ell_\mathrm{cc}}>\qty{100}{nm}$. Orange represents the hydrodynamic regime where $\min{\ell_\mathrm{cr}}>\qty{2}{\um}$ and $\max{\ell_\mathrm{cc}}<\qty{100}{nm}$. Green represents the diffusive regime where $\max{\ell_\mathrm{cr}}<\qty{2}{\um}$. Different shadings correspond to thresholds of normalized RMSD~$<1.1,1.2,1.3$ and gray regions indicate regime crossovers. Dashed lines delimit the extent of the regime of fluctuating magnetic moments and negative $\mathrm{d}R/\mathrm{d}T$ reported in \cite{Holleis_Fluctuating_2025}. 
\textbf{b}, Phase diagram with a diffusive boundary using alternative thresholds $\ell_\mathrm{cr}=\qty{1}{\um}$ and $\ell_\mathrm{cc}=\qty{200}{nm}$. 
Compared to panel a, here the hydrodynamic regime expands at the expense of ballistic and diffusive regimes
\textbf{c}, Phase diagram with a mixed (half-diffusive, half-specular) boundary using thresholds $\ell_\mathrm{cr}=\qty{4}{\um}$ and $\ell_\mathrm{cc}=\qty{50}{nm}$. 
\textbf{d}, Phase diagram with a mixed boundary using $\ell_\mathrm{cr}=\qty{2}{\um}$ and $\ell_\mathrm{cc}=\qty{100}{nm}$. 
For the same thresholds, the mixed boundary results in a larger hydrodynamic region than the diffusive boundary, consistent with the observations presented in Extended Data Fig.~\ref{fig:ED_boundary}. 
}
\label{fig:ED_alt_phase}
\end{figure*}

\begin{figure}[ht!]
\centering
\includegraphics[width=3.5in]{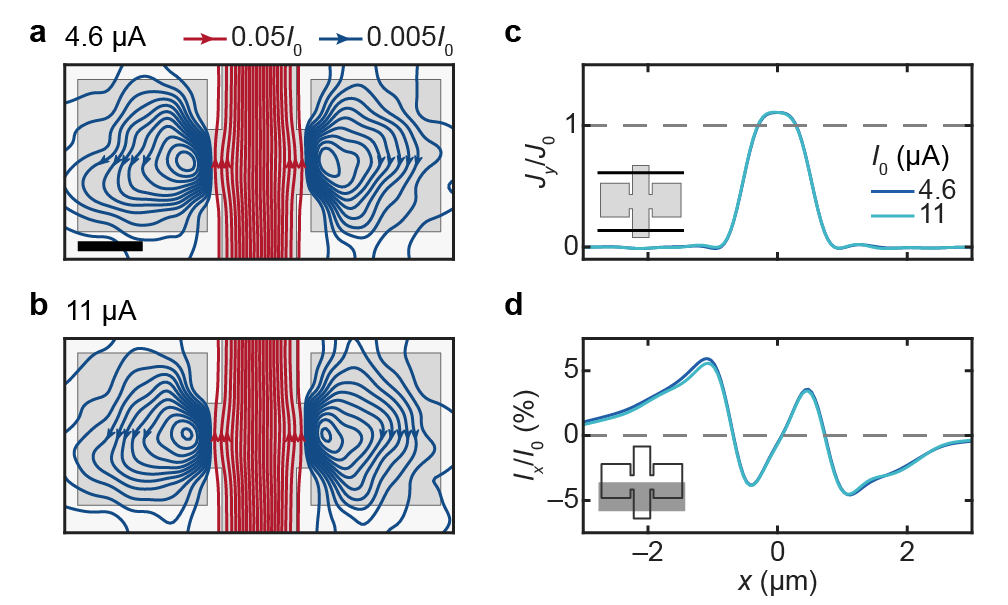}
\caption{\textbf{Linearity under low current drive.}
\textbf{a}, Current streamlines for $n_\mathrm{e}=\qty{-1.5e12}{cm^{-2}}$ and $D=\qty{0.39}{V/nm}$ for $I_0=\qty{4.6}{\uA}$, and 
\textbf{b}, $I_0=\qty{11}{\uA}$. Scale bar: \qty{1}{\um}. 
\textbf{c}, $J_y(x,\pm\qty{2}{\um})/J_0$, and 
\textbf{d}, $I_x(x)/I_0$ for the data in panels a and b. 
}
\label{fig:ED_linearity}
\end{figure}

\begin{figure}[ht!]
\centering
\includegraphics[width=3.5in]{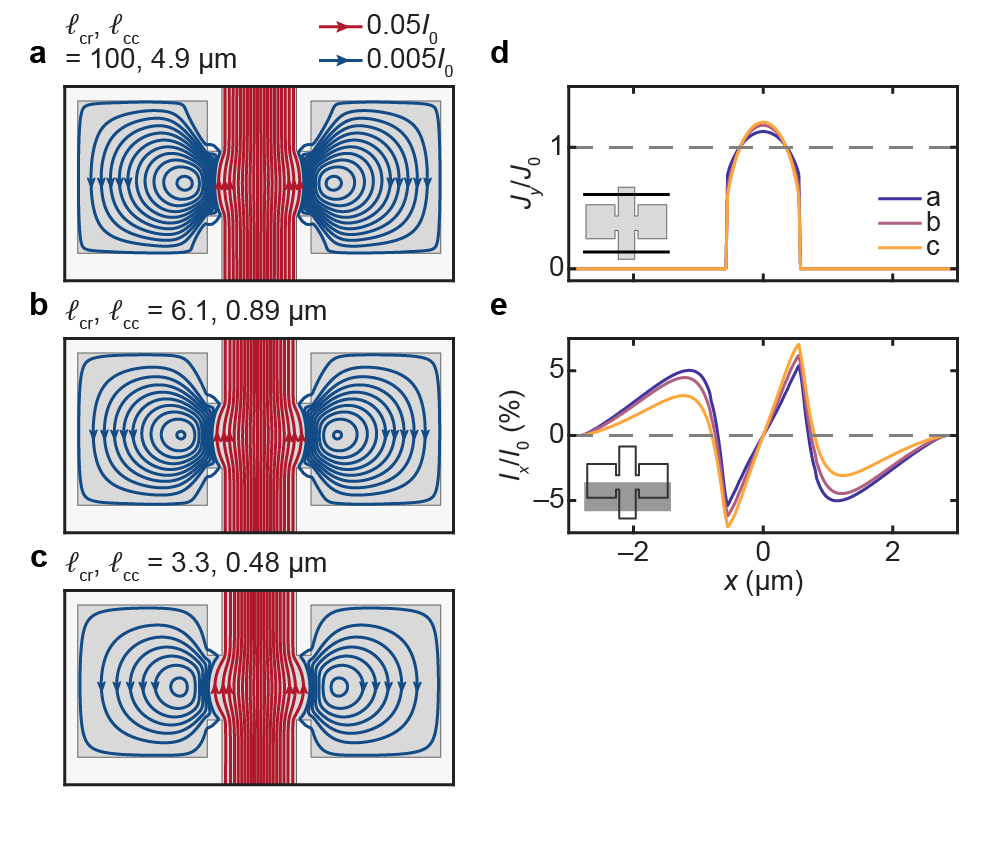}
\caption{\textbf{Simulating current-driven ballistic-to-viscous crossover.}
\textbf{a}, Theoretical current streamlines for $\ell_\mathrm{cr}=\qty{100}{\um}$ and $\ell_\mathrm{cc}=\qty{4.9}{\um}$, corresponding to the best fit for the data of Fig.~\ref{fig:fig4}a. Scale bar: \qty{1}{\um}.
\textbf{b}, Theoretical streamlines for $\ell_\mathrm{cr}=\qty{6.1}{\um}$ and $\ell_\mathrm{cc}=\qty{890}{nm}$, corresponding to the best fit for the data of Fig.~\ref{fig:fig4}b. 
\textbf{c}, Theoretical streamlines for $\ell_\mathrm{cr}=\qty{3.3}{\um}$ and $\ell_\mathrm{cc}=\qty{480}{nm}$, corresponding to the best fit for the data of Fig.~\ref{fig:fig4}c. 
\textbf{d}, $J_y(x,\pm\qty{2}{\um})/J_0$ corresponding to panels a, b, and c. 
\textbf{e}, $I_x(x)/I_0$ corresponding to panels a, b, and c. 
The increase in channel flow concentration and suppression of vortical flow upon decreasing $\ell_\mathrm{cc}$ reproduce the observed current-driven behavior. 
}
\label{fig:ED_fit_Fig4}
\end{figure}

\begin{figure}[ht!]
\centering
\includegraphics[width=3in]{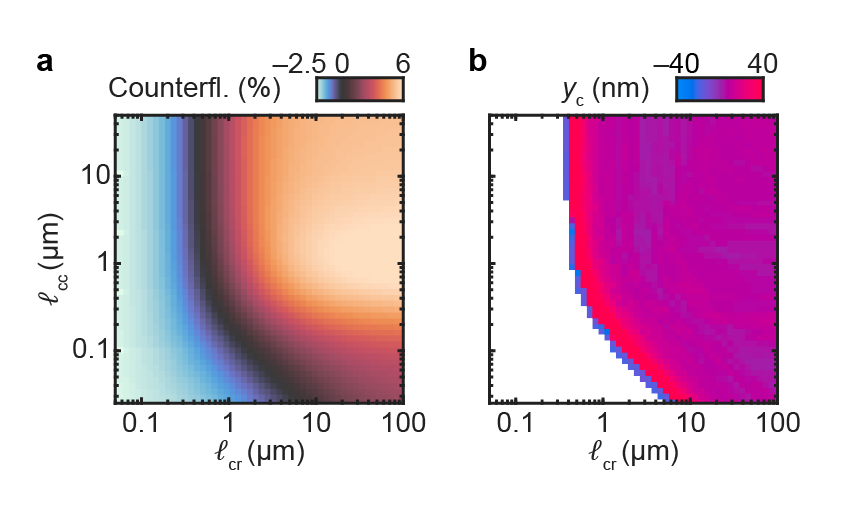}
\caption{\textbf{Vortex center position within the linearized Boltzmann model.}
\textbf{a}, Counterflow $I_x(|x|=\qty{1.3}{\um})/I_0$ within the Boltzmann model parameterized by $\ell_\mathrm{cr}$ and $\ell_\mathrm{cc}$. 
\textbf{b}, The extracted vortex center position $y_\mathrm{c}$ (see Methods). No vortices are present in the diffusive regime (white region and corresponding to the area of negative counterflow in panel a). 
}
\label{fig:ED_center}
\end{figure}

\begin{figure*}[ht!]
\centering
\includegraphics[width=7in]{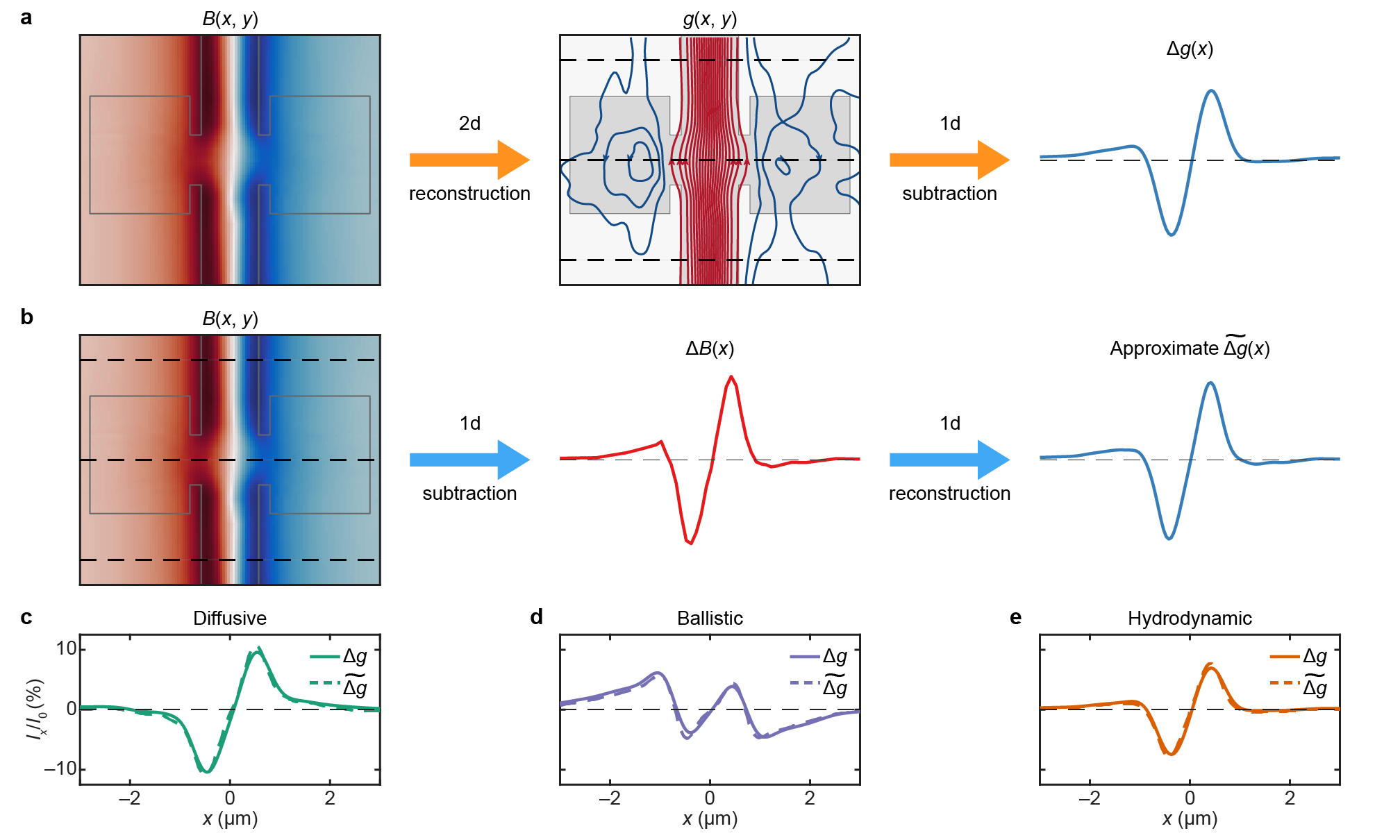}
\caption{\textbf{Comparison of full 2d current reconstruction and simplified 1d reconstruction.}
\textbf{a}, 2d $g(x,y)$ map is inverted from $B(x,y)$ using the full kernel \eqref{eq:g2B}, and $\Delta g(x)$ is obtained by subtracting it along three 1d trajectories (Eq.~\ref{eq:IxDg}). 
\textbf{b}, Approximate $\tilde{\Delta g}(x)$ is directly inverted from $\Delta B(x)$ using the effective 1d kernel \eqref{eq:g2B_1d} (where we set $l_y=\qty{3}{\um}$). 
\textbf{c}, Comparison between normalized $\Delta g(x)$ and $\tilde{\Delta g}(x)$ (i.e., $I(x)/I_0$) at $n_\mathrm{e}=\qty{0e12}{cm^{-2}}$ and $D=\qty{0}{V/nm}$ (diffusive regime). 
\textbf{d}, Comparison between $\Delta g(x)$ and $\tilde{\Delta g}(x)$ at $n_\mathrm{e}=\qty{-1.3e12}{cm^{-2}}$ and $D=\qty{0.39}{V/nm}$ (ballistic regime). 
\textbf{e}, Comparison between $\Delta g(x)$ and $\tilde{\Delta g}(x)$ at $n_\mathrm{e}=\qty{-0.4e12}{cm^{-2}}$ and $D=\qty{0.39}{V/nm}$ (hydrodynamic regime). 
}
\label{fig:ED_rec_2d1d}
\end{figure*}

% Extended Data Video
\setcounter{figure}{0}
\renewcommand{\figurename}{\textbf{Supplementary Video}}

\begin{figure*}[ht!]
\centering
\caption{\textbf{Current-driven vortex deformation.}
Evolution of current streamlines at $n_\mathrm{e}=\qty{-0.25e12}{cm^{-2}}$ and $D=\qty{0}{V/nm}$ upon increasing $I_0$ from 27 to \qty{100}{\uA}. Scale bar: \qty{1}{\um}. 1 red line = $0.05I_0$, 1 solid blue line = $0.005I_0$. Additional dashed streamlines serve as guides to the eye. 
}
\label{video1}
\end{figure*}

\end{document}